\documentclass[longauth]{aa}

\usepackage{graphicx}
\usepackage{txfonts}
\usepackage{gensymb} 
\usepackage{xspace}
\usepackage{tabularx}

\usepackage{natbib,twoopt}
\usepackage[breaklinks=true]{hyperref} 
\bibpunct{(}{)}{;}{a}{}{,} 
\makeatletter
\newcommandtwoopt{\citeads}[3][][]{\href{http://adsabs.harvard.edu/abs/#3}%
{\def\hyper@linkstart##1##2{}%
\let\hyper@linkend\@empty\citealp[#1][#2]{#3}}}
\newcommandtwoopt{\citepads}[3][][]{\href{http://adsabs.harvard.edu/abs/#3}%
{\def\hyper@linkstart##1##2{}%
\let\hyper@linkend\@empty\citep[#1][#2]{#3}}}
\newcommandtwoopt{\citetads}[3][][]{\href{http://adsabs.harvard.edu/abs/#3}%
{\def\hyper@linkstart##1##2{}%
\let\hyper@linkend\@empty\citet[#1][#2]{#3}}}
\newcommandtwoopt{\citeyearads}[3][][]%
{\href{http://adsabs.harvard.edu/abs/#3}
{\def\hyper@linkstart##1##2{}%
\let\hyper@linkend\@empty\citeyear[#1][#2]{#3}}}
\makeatother
\def\ms{\hbox{m\,s$^{-1}$}}         
\def\m2s2{\hbox{\,m$^{2}$\,s$^{-2}$}} 
\def\kms{\hbox{\,km\,s$^{-1}$}}       
\def\vsini{\hbox{$v$\,sin\,$i_{\star}$}}      
\def\Msun{\hbox{$M_{\odot}$}}             
\def\Rsun{\hbox{$R_{\odot}$}}

\def\ten[#1]{$\;\times 10^{#1}$}

\def\logg{$\log g$}

\newcommand{\e}[1]{{\times10^{#1}}}

\newcommand{\Rnom}{\hbox{$\mathcal{R}^{\rm N}_{\odot}$}} 

\newcommand{\GMnom}{\hbox{$\mathcal{(GM)}^{\rm N}_{\odot}$}}

\newcommand{\Renom}{\hbox{$\mathcal{R}^{\rm N}_{e \rm E}$}}

\newcommand{\GMenom}{\hbox{$\mathcal{(GM)}^{\rm N}_{\rm E}$}}

\newcommand{\rebound}{{\sc \tt REBOUND}\xspace}
\newcommand{\whf}{{\sc \tt WHFast}\xspace}
\newcommand{\emcee}{{\sc \tt emcee}\xspace}
\newcommand{\juliet}{{\sc \tt juliet}\xspace}
\newcommand{\batman}{{\sc \tt batman}\xspace}
\newcommand{\radvel}{{\sc \tt radvel}\xspace}
\newcommand{\celerite}{{\sc \tt celerite}\xspace}
\newcommand{\george}{{\sc \tt george}\xspace}
\newcommand{\dynesty}{{\sc \tt dynesty}\xspace}

\newcommand{\specmatch}{{\sc \tt SpecMatch-Emp}\xspace}

\newcommand{\REarth}{$\mathrm{R_E}$\xspace}
\newcommand{\MEarth}{$\mathrm{M_E}$\xspace}

\begin{document} 

     \title{GJ 3090 b: one of the most favourable mini-Neptune for atmospheric characterisation}

   \author{J.M.~Almenara\inst{\ref{grenoble}}
        \and X.~Bonfils\inst{\ref{grenoble}}
        \and J.F.~Otegi\inst{\ref{geneva},\ref{zurich}}
        \and M.~Attia\inst{\ref{geneva}}
        \and M.~Turbet\inst{\ref{geneva}}
        \and N.~Astudillo-Defru\inst{\ref{concepcion}}
        \and K.A.~Collins\inst{\ref{harvard}}
        \and A.S.~Polanski\inst{\ref{kansas}}
        \and V.~Bourrier\inst{\ref{geneva}}
        \and C.~Hellier\inst{\ref{keele}}
        \and C.~Ziegler\inst{\ref{texas}}
        \and F.~Bouchy\inst{\ref{geneva}}
        \and C.~Brice\~{n}o\inst{\ref{tololo}}
        \and D.~Charbonneau\inst{\ref{harvard}}
        \and M.~Cointepas\inst{\ref{grenoble},\ref{geneva}}
        \and K.I.~Collins\inst{\ref{gmu}} 
        \and I.~Crossfield\inst{\ref{kansas}}   
        \and X.~Delfosse\inst{\ref{grenoble}}
        \and R.F.~D\'{i}az\inst{\ref{ba}}
        \and C.~Dorn\inst{\ref{zurich}}
        \and J.P.~Doty\inst{\ref{noqsi}} 
        \and T.~Forveille\inst{\ref{grenoble}}
        \and G.~Gaisn\'{e}\inst{\ref{grenoble}}
        \and T.~Gan\inst{\ref{tsinghua}} 
        \and R.~Helled\inst{\ref{zurich}}
        \and K.~Hesse\inst{\ref{mit}}
        \and J.M.~Jenkins\inst{\ref{ames}}
        \and E.L.N.~Jensen\inst{\ref{swarthmore}} 
        \and D.W.~Latham\inst{\ref{harvard}}
        \and N.~Law\inst{\ref{carolina}}
        \and A.W.~Mann\inst{\ref{carolina}}
        \and S.~Mao\inst{\ref{tsinghua}}
        \and B.~McLean\inst{\ref{stsi}}
        \and F.~Murgas\inst{\ref{iac},\ref{ull}} 
        \and G.~Myers\inst{\ref{myers}} 
        \and S.~Seager\inst{\ref{mit},\ref{mit2},\ref{mit3}}
        \and A.~Shporer\inst{\ref{mit}}
        \and T.G.~Tan\inst{\ref{pest},\ref{curtin}} 
        \and J.D.~Twicken\inst{\ref{seti},\ref{ames}}
        \and J.~Winn\inst{\ref{princeton}}
      }

      \institute{
        Univ. Grenoble Alpes, CNRS, IPAG, F-38000 Grenoble, France\label{grenoble}
        \and Observatoire de Gen\`eve, Département d’Astronomie, Universit\'e de Gen\`eve, Chemin Pegasi 51, 1290 Versoix, Switzerland\label{geneva}
        \and Institute for Computational Science, University of Zurich, Winterthurerstr. 190, CH-8057 Zurich, Switzerland \label{zurich}
        \and Departamento de Matem\'{a}tica y F\'{i}sica Aplicadas, Universidad Cat\'{o}lica de la Sant\'{i}sima Concepci\'{o}n, Alonso de Rivera 2850, Concepci\'{o}n, Chile\label{concepcion}
        \and Center for Astrophysics \textbar \ Harvard \& Smithsonian, 60 Garden Street, Cambridge, MA 02138, USA\label{harvard}
        \and Department of Physics and Astronomy, University of Kansas, Lawrence, KS 66045\label{kansas}
        \and Astrophysics Group, Keele University, Staffordshire ST5 5BG, UK\label{keele}
        \and Department of Physics, Engineering and Astronomy, Stephen F. Austin State University, 1936 North St, Nacogdoches, TX 75962, USA\label{texas}
        \and SOAR Telescope/NSF’s NOIRLab, Casilla 603, La Serena, Chile\label{tololo}
        \and George Mason University, 4400 University Drive, Fairfax, VA, 22030 USA\label{gmu}
        \and International Center for Advanced Studies (ICAS) and ICIFI (CONICET), ECyT-UNSAM, Campus Miguelete, 25 de Mayo y Francia, (1650) Buenos Aires, Argentina\label{ba}
        \and Noqsi Aerospace Ltd., 15 Blanchard Avenue, Billerica, MA 01821, USA\label{noqsi}
        \and Department of Astronomy and Tsinghua Centre for Astrophysics, Tsinghua University, Beijing 100084, China\label{tsinghua}
        \and Department of Physics and Kavli Institute for Astrophysics and Space Research, Massachusetts Institute of Technology, Cambridge, MA 02139, USA\label{mit}
        \and NASA Ames Research Center, Moffett Field, CA 94035, USA\label{ames}
        \and Dept.\ of Physics \& Astronomy, Swarthmore College, Swarthmore PA 19081, USA\label{swarthmore}
        \and Department of Physics and Astronomy, The University of North Carolina at Chapel Hill, Chapel Hill, NC 27599-3255, USA\label{carolina}
        \and Space Telescope Science Institute, 3700 San Martin Drive, Baltimore, MD, 21218, USA\label{stsi}
        \and Instituto de Astrofísica de Canarias (IAC), E-38200 La Laguna, Tenerife, Spain\label{iac}
        \and Dept. Astrofísica, Universidad de La Laguna (ULL), E-38206 La Laguna, Tenerife, Spain\label{ull}
        \and AAVSO, 5 Inverness Way, Hillsborough, CA 94010, USA\label{myers}
        \and Department of Earth, Atmospheric and Planetary Sciences, Massachusetts Institute of Technology, Cambridge, MA 02139, USA\label{mit2}
        \and Department of Aeronautics and Astronautics, MIT, 77 Massachusetts Avenue, Cambridge, MA 02139, USA\label{mit3}
        \and Perth Exoplanet Survey Telescope, Perth, Western Australia\label{pest}
        \and Curtin Institute of Radio Astronomy, Curtin University, Bentley, Western Australia 6102\label{curtin}
        \and SETI Institute, Mountain View, CA 94043, USA\label{seti}
        \and Department of Astrophysical Sciences, Princeton University, NJ 08544, USA\label{princeton}
        }
      \date{}

      \date{}

 
  \abstract
      {
        We report the detection of GJ~3090~b (TOI-177.01), a mini-Neptune on a 2.9-day orbit transiting a bright (K = 7.3~mag) M2 dwarf located at 22~pc. The planet was identified by the Transiting Exoplanet Survey Satellite and was confirmed with the High Accuracy Radial velocity Planet Searcher radial velocities. Seeing-limited photometry and speckle imaging rule out nearby eclipsing binaries. Additional transits were observed with the LCOGT, Spitzer, and ExTrA telescopes. We characterise the star to have a mass of 0.519~$\pm$~0.013~\Msun\ and a radius of 0.516~$\pm$~0.016~\Rsun. We modelled the transit light curves and radial velocity measurements and obtained a planetary mass of 3.34~$\pm$~0.72~\MEarth, a radius of 2.13~$\pm$~0.11~\REarth, and a mean density of 1.89$^{+0.52}_{-0.45}~{\rm g/cm^3}$. The low density of the planet implies the presence of volatiles, and its radius and insolation place it immediately above the radius valley at the lower end of the mini-Neptune cluster. A coupled atmospheric and dynamical evolution analysis of the planet is inconsistent with a pure H-He atmosphere and favours a heavy mean molecular weight atmosphere. 
        The transmission spectroscopy metric of 221$^{+66}_{-46}$ means that GJ~3090~b is the second or third most favourable mini-Neptune after GJ~1214~b whose atmosphere may be characterised. At almost half the mass of GJ~1214~b, GJ~3090~b is an excellent probe of the edge of the transition between super-Earths and mini-Neptunes.   
        We identify an additional signal in the radial velocity data that we attribute to a planet candidate with an orbital period of 13~days and a mass of 17.1$^{+8.9}_{-3.2}$~\MEarth, whose transits are not detected. 
      }
   \authorrunning{J.M. Almenara et al.}
   \titlerunning{GJ~3090}

   \maketitle
%

\section{Introduction}\label{section.introduction}

The structure, bulk composition, and atmosphere of planets that are found in transit can be characterised. To achieve precise measurements, planets that transit bright and small stars are especially well suited. 

Over the past two decades, $\sim$3{,}500 transiting planets have been discovered\footnote{according to \url{https://exoplanetarchive.ipac.caltech.edu} as of 2021, October 12th} and important statistical properties have emerged. Super-Earths and mini-Neptunes, which are absent in our Solar System, have been found to be frequent around other stars for a wide range of orbital periods \citep{howard2012}. After the seminal discovery of HD209458~b \citep{charbonneau2000,henry2000}, many other planets have been detected on short-period orbits with transit surveys ($P<10$~days), but a relative scarcity is observed for hot Neptunes; this has been called the 'hot-Neptune desert' \citep{lecavelier2007,davis2009,szabo2011,beauge2013,lundkvist2016,mazeh2016}. Another robust outcome of recent research is the bimodal radius distribution of planets smaller than $\sim 4$~\REarth. The two modes seem to divide rocky from volatile-rich planets with a transition at around $1.8$~\REarth \citep{fulton2017,zeng2019,mousis2020} that may depend on orbital period \citep{vaneylen2018,petigura2020} and stellar mass \citep{cloutier2020}. The wish to interpret these demographics has inspired much theoretical work. Models of photoevaporation and orbital evolution have been shown to reproduce the radius valley \citep{owen2017,jin2018}. Evaporation can also be powered by the intrinsic heat of the planet that accumulates during the phase of accretion, which might play a role for small exoplanets and for the formation of their bimodal distribution \citep{ginzburg2018,gupta2019}. Today, discovering new planets near the edge of the valley and around stellar hosts of various masses is particularly important for understanding where these mechanisms are predominant. In particular, the precise characterisation of the atmospheric composition of mini-Neptunes can reveal clues for their formation location and migration and for their interior structure \citep{bean2021}. Mini-Neptunes around M dwarfs are generally favourable targets for atmospheric characterisation through transit spectroscopy because of their large-scale height and the ratio of the planet-to-star radius, although the final amplitude of the signal also depends on the atmospheric composition and the presence of clouds or hazes \citep{Kreidberg2014}. 

We report here the confirmation of GJ~3090~b, a mini-Neptune around a bright M dwarf (GJ\,3090, HIP\,6365) whose atmosphere can be characterised. GJ~3090~b was first identified by the Transiting Exoplanet Survey Satellite \citep[TESS;][]{ricker2015}, which is a space-based all-sky transit survey mission that focuses on bright targets.

In Sect.~\ref{section.observations} we describe the data we used to detect the candidate GJ~3090~b and confirm its planetary nature. In Sect.~\ref{section.stellar_parameters} we characterise its host star, and in Sect.~\ref{section.analysis} we model the radial velocity (RV) and transit data. Finally, in Sect.~\ref{section.results} we discuss the results of our work, including interior characterisation, coupled atmospheric and dynamical evolution, and prospects for atmospheric characterisation.

\section{Observations}\label{section.observations}

\subsection{TESS}\label{sec.tess}

TESS observed GJ~3090 (TIC 262530407) in sectors 2, 3, and 29 with a two-minute cadence for a total of 81~days and a time span of 760~days. A transit signature with a 2.853-day period and 1.7~mmag depth was first identified in the Science Processing Operations Center \citep[SPOC;][]{jenkins2016} transiting planet search \citep{jenkins2002,jenkins2010,jenkins2020} of the sector~2 light curve. The SPOC threshold crossing event was subsequently promoted by the TESS Science Office to TESS object of interest \citep[TOI;][]{guerrero2021} status as TOI 177.01 based on the model fit and diagnostic test results in the SPOC data validation report \citep{twicken2018,li2019}.

\begin{figure*}
    \centering  
    \includegraphics[width=1.0\textwidth]{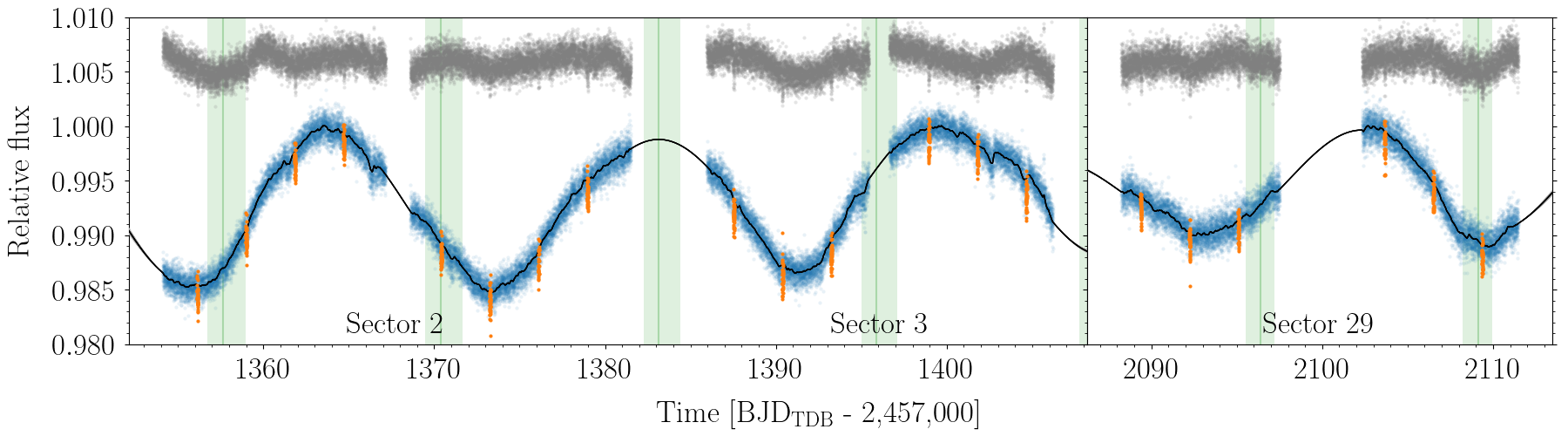} 
    \caption{TESS-normalised SAP (blue points) and PDCSAP (grey points, offset for clarity). Transits of GJ~3090~b (orange) were excluded before fitting the data with a QP kernel GP (the black line is the posterior median model). The vertical green lines and bands are the median and 68.3\% credible interval transit window for a second planet candidate in the system (Sects.~\ref{sec.RVanalysis} and \ref{sec.juliet}).}
\label{fig.tess}
\end{figure*}

The SPOC simple aperture photometry \citep[SAP;][]{twicken2010,morris2020} and pre-search data conditioning SAP \citep[PDCSAP,][]{strumpe2012,strumpe2014,smith2012} light curves are shown in Fig.~\ref{fig.tess}. The PDCSAP light curve is the SAP light curve from which long-term trends have been removed. Twenty complete transits were observed. The SAP light curve shows an amplitude variability (peak to peak) of $\sim$1.5\% that is compatible with the stellar rotational modulation. To estimate the stellar rotation period, we removed the transits from the SAP light curve, normalised each sector by the maximum flux of a 0.5-day sliding mean, and modelled the resulting light curve with \juliet \citep{espinoza2019, speagle2020} and the quasi-periodic (QP) kernel Gaussian process (GP) included in \celerite \citep{foreman-mackey2017}. We obtained a period of the QP kernel of 17.65~$\pm$~0.48~days (Fig.~\ref{fig.tess}).

Next, we prepared the TESS data for transit modelling as follows: We started from the PDCSAP light curve, that include corrections for contamination in the aperture of 1.5, 1.3, and 2.0\% for sectors 2, 3, and 29 respectively, mainly due to one contaminant at 28\arcsec\ that is $\sim$4~magnitudes fainter than the target in the Gaia G band (Fig.~\ref{fig.tpfplotter}). We extracted the data spanning three transit durations around the centre of each transit. Finally, we normalised each transit with a straight line fitted to the out-of-transit parts of the light curve. To account for the differences in stellar flux levels at the times of each transit \citep{czesla2009}, we corrected the normalised transit curves using the GP model described above. To properly correct for the stellar flux variability, it would be necessary to have observed the purely photospheric stellar surface without any activity regions. This does not seem to be the case here, as there is no global flat flux maximum. In any case, this is a minor correction that changes the planet-to-star radius ratio by only $\sim$0.2~$\sigma$.

\subsection{Seeing-limited photometry}

We conducted ground-based photometric follow-up observations of GJ~3090 as part of the {\em TESS} follow-up observing program\footnote{\url{https://tess.mit.edu/followup}} \citep[TFOP;][]{collins:2019} to attempt to rule out or identify nearby eclipsing binaries (NEBs) as potential sources of the \textit{TESS} detection, measure the transit-like event on target to confirm the depth and thus the \textit{TESS} photometric deblending factor, and refine the \textit{TESS} ephemeris. We used the {\tt TESS Transit Finder}, which is a customised version of the {\tt Tapir} software package \citep{Jensen:2013}, to schedule our transit observations. 

The observations are summarised in Table \ref{table:tfopobservatories}. Three observations searched for and found no NEBs within $2.5\arcmin$ of the target star. One observation was inconclusive, and two observations detected the event on-target, as indicated in Table \ref{table:tfopobservatories}. The Las Cumbres Observatory Global Telescope \citep[LCOGT;][]{Brown:2013} observations were calibrated with the standard LCOGT {\tt BANZAI} pipeline \citep{McCully:2018}. All photometric data were extracted using {\tt AstroImageJ} \citep{Collins:2017}, except for the Perth Exoplanet Survey Telescope observation, which used a custom pipeline based on {\tt C-Munipack}\footnote{\url{http://c-munipack.sourceforge.net}}.

\begin{table*}[h]
\caption{TFOP photometric follow-up observations.}\label{table:tfopobservatories}
\begin{tabular}{lccccc}
  \hline
  \hline  
Observatory / Location & UTC Date & Aperture (m) & Filter & Coverage & Result \\
\hline
Perth Exoplanet Survey Telescope, Perth, Australia & 2018-10-24 & 0.3   &   Rc   & Full     & No NEB detected \\
Myers Observatory, Siding Spring, Australia        & 2018-10-24 & 0.4   &   Rc   & Full     & No NEB detected\\
LCOGT CTIO, Chile & 2018-11-01 & 0.4 &   $i'$ & Full     & Inconclusive\\
LCOGT SAAO, South Africa & 2018-11-07 & 0.4 & $r'$ & Full     & On-target detection\\
Myers Observatory, Siding Spring, Australia        & 2018-11-10 & 0.4   &   Lum  & Full     & No NEB detected\\
LCOGT CTIO, Chile & 2018-11-19 & 1.0 &   $r'$ & Ingress  & On-target detection\\
\hline
\end{tabular}
\tablefoot{The filter designation Lum indicates a luminance filter with central wavelength 550\,nm and bandwidth 300\,nm. LCOGT indicates observations from the Las Cumbres Observatory Global Telescope \citep[][]{Brown:2013} 1.0\,m and 0.4\,m network nodes at the Cerro Tololo Inter-American Observatory (CTIO) and at the South Africa Astronomical Observatory (SAAO).}
\end{table*}

\subsection{Spitzer}\label{sec.spitzer}

GJ~3090 was observed with \textit{Spitzer} on 2019 April 4, using director's discretionary time \citep{Crossfield2019}. A single transit was observed using the 4.5~$\mu$m channel \citep[IRAC2;][]{Fazio2005} in subarray mode with an integration time of 2~seconds. The transit observation spanned 18 hr 37 min, totalling 711 frames with short observations taken before and after transit to check for bad pixels. Peak-up mode was used to place the star as close as possible to the well-characterised sweet spot of the detector.

\subsection{ExTrA}

The ExTrA facility \citep{bonfils2015}, located at La Silla observatory, consists of a near-infrared (0.85 to 1.55~$\mu$m) multi-object spectrograph fed by three 60 cm telescopes. Five fibre positioners intercept the light from a target and four comparison stars at the focal plane of each telescope. We observed three transits of GJ~3090~b on nights UTC 2019 December 14, 2020 January 3, and 2020 November 29. On the first and second nights, we observed simultaneously with two telescopes, whereas on the third night, we observed with three telescopes. We used the fibres with 8\arcsec\ apertures, the higher resolution mode of the spectrograph (R$\sim$200), and 60-second exposures. The comparison stars were 2MASS J01235337-4659196, 2MASS J01235446-4647126, 2MASS J01241189-4633588, and 2MASS J01224245-4617544. The ExTrA data were analysed using custom data reduction software.

\subsection{WASP}\label{sec.wasp}

\begin{figure}
    \centering  
    \includegraphics[width=0.48\textwidth]{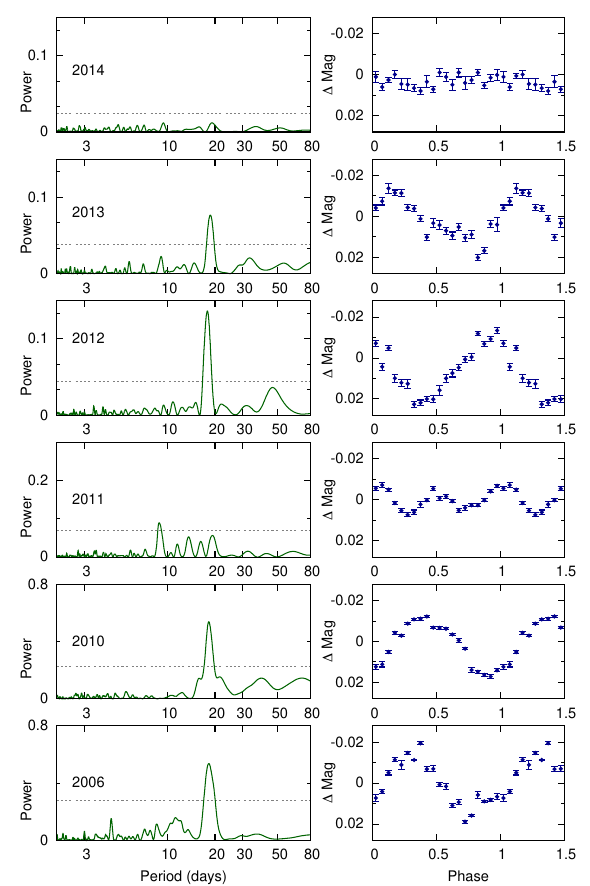} 
    \caption{Periodograms of the WASP-South data for GJ~3090 (left) and the resulting folds on the 18.2 d rotational period (right). The dotted lines mark the estimated 1\% likelihood false-alarm level.}
\label{fig.wasp}
\end{figure}

WASP-South was a wide-field array of eight cameras forming the southern station of the WASP transit-search survey \citep{pollacco2006}. The field of GJ~3090 was observed over a span of 200 nights in 2006, and then over 170-night spans in 2010 and 2011. During that time, WASP-South was equipped with 200 mm f/1.8 lenses, observing with a 400--700 nm passband, and with a photometric extraction aperture of 48 arcsec. GJ~3090 was also observed over 180-night spans in  2012, 2013, and 2014 each, when WASP-South was equipped with 85-mm f/1.2 lenses using an SDSS-$r$ filter. In total, 71\,000 photometric observations were obtained, with a typical cadence of 12 minutes. GJ~3090 is the brightest star in the extraction apertures. The next brightest star is 4 magnitudes fainter and also dominates the TESS contamination correction. We searched each season of data for modulations due to stellar rotation using the methods from \citet{maxted2011}. We show the results in Fig.~\ref{fig.wasp}. 

A clear rotational modulation is seen at a period of 18.2~$\pm$~0.4~days. The error is estimated from the dispersion between different datasets and is dominated by phase shifts of the modulation. In most years, the modulation has an amplitude of 12 to 18 mmag. In 2011, the modulation is weaker and the dominant power is at the first harmonic, and in 2014, the modulation is below our detection threshold. 

\subsection{SOAR speckle imaging}

\begin{figure}
  \centering
  \includegraphics[width=0.47\textwidth]{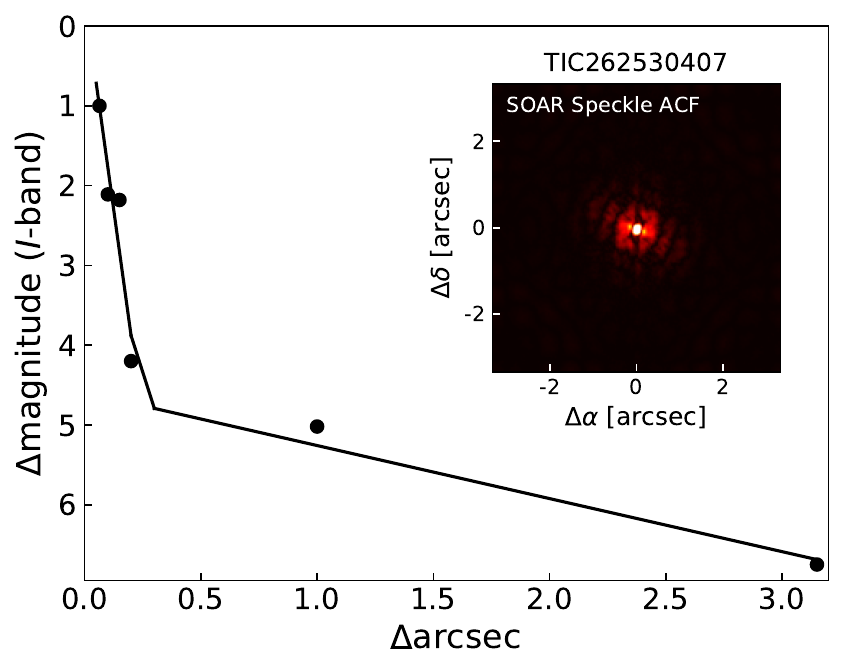}
  \caption{I-band 5$\sigma$ contrast curve from SOAR speckle imaging of GJ~3090 (TIC 262530407). The inset depicts the corresponding speckle autocorrelation function.} \label{fig.soar}
\end{figure}

High-angular resolution imaging can identify nearby sources that contaminate the TESS photometry, resulting in an underestimated planetary radius or in  astrophysical false positives such as background eclipsing binaries. We searched for stellar companions to GJ~3090 with speckle imaging on the 4.1 m Southern Astrophysical Research (SOAR) telescope \citep{tokovinin2018} on UT 2018 October 21, observing in Cousins I band, similar to the TESS bandpass. More details of the observation are available in \citet{ziegler2020}. The 5$\sigma$ detection sensitivity and speckle autocorrelation functions from the observations are shown in Fig.~\ref{fig.soar}. The SOAR observations detect no additional stars within 3\arcsec\ of GJ~3090.

\subsection{HARPS}\label{sec.harps}

We obtained differential RVs of GJ~3090 with HARPS \citep{mayor2003}, the optical velocimeter installed at the ESO~3.6~m telescope at La Silla Observatory. Under the ESO program 1102.C-0339(A), we collected 55~spectra over a time span of 326~days, with an exposure time of 900~seconds each (one spectrum taken on 2019 January 24 has a 1800-second exposure time), slow readout mode (104~kHz), and without simultaneous wavelength calibration. The signal-to-noise ratio at 650~nm ranges between 20 and 71, with a mean value of 42. The spectra were reduced with the HARPS Data Reduction Software\footnote{\url{http://www.eso.org/sci/facilities/lasilla/instruments/harps/doc/DRS.pdf}}. We used the template-matching algorithm described in \citet{Astudillo-Defru2017b} to compute the RVs, which have a median RV precision of 2.0~m/s (dominated by photon noise) and a dispersion of 8.5~m/s. The RVs, cross-correlation function (CCF) characteristics, and activity indicators are listed in Table~\ref{table.rv}.
We measured $\log R'_{HK} = -4.416 \pm 0.048$ in the average of the HARPS spectra, from which the \cite{astudillo2017} activity-rotation relation estimates a rotation period of $P_{\rm rot} = 17.0 \pm 1.3$~days that excellently agrees with the results presented in Sects.~\ref{sec.tess} and \ref{sec.wasp}.

\section{Stellar parameters}\label{section.stellar_parameters}

\begin{figure}
  \centering
  \includegraphics[width=0.49\textwidth]{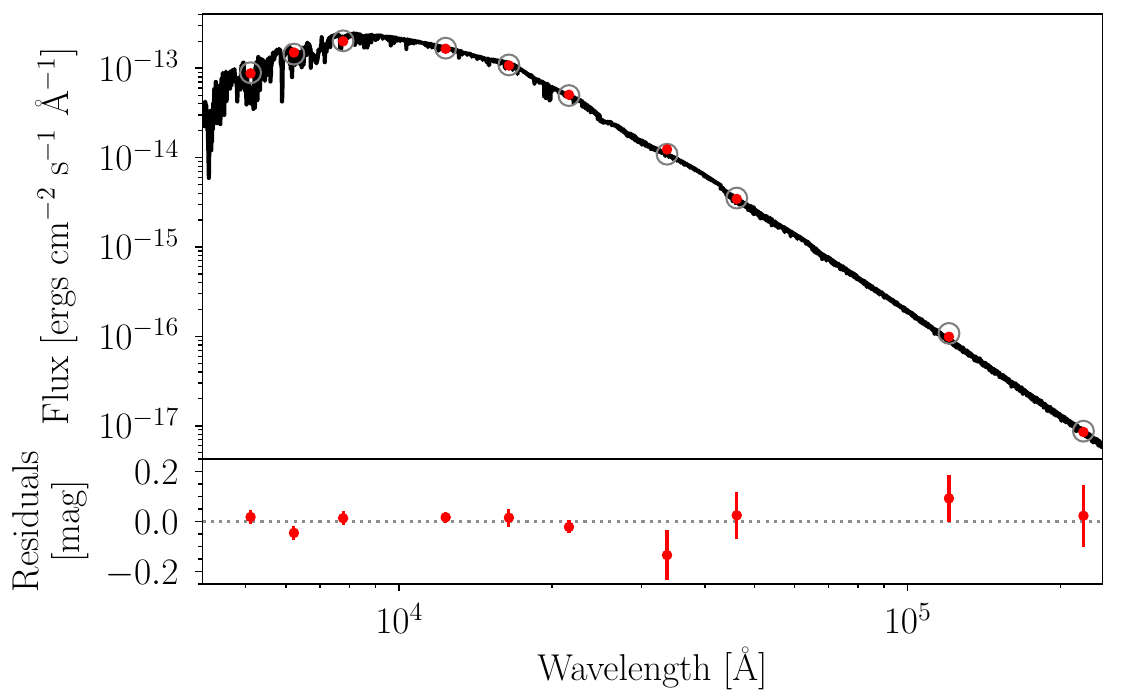}
  \caption{SED of GJ~3090. Top panel: MAP PHOENIX/BT-Settl interpolated synthetic spectrum (solid line). Red circles show photometric observations, and grey open circles are the result of integrating the synthetic spectrum in the observed bandpasses. Bottom panel: Residuals of the MAP model (the jitter has been added quadratically to the error bars of the data).} \label{fig.sed}
\end{figure}

We derived the mass and radius of GJ~3090 from empirical relations based on luminosity. We used the Gaia-corrected \citep{gaia2016,gaiaEDR3,lindegren2021} parallax determination (44.555~$\pm$~0.027~mas) to compute the distance (22.444~$\pm$~0.013~pc) and an absolute magnitude of $M_{K_s}~=~5.538~\pm~0.026$. We then used the empirical relations of \citet{mann2019} and \citet{mann2015} to derive a mass of $M_{\star}~=~0.519~\pm~0.013~\Msun$ and a radius of $R_{\star}~=~0.517~\pm~0.016~\Rsun$, respectively, which were used as priors for Sect.~\ref{sec.photodynamical}. We derived the stellar radius independently with the spectral energy distribution (SED) that we constructed using the magnitudes from Gaia \citep{riello2021}, the 2-Micron All-Sky Survey \citep[2MASS;][]{2mass,cutri2003}, and the Wide-field Infrared Survey Explorer \citep[WISE;][]{wise,cutri2013}. The measurements are listed in Table~\ref{table.stellar_parameters}. We modelled these magnitude measurements using the procedure described in \citet{diaz2014}, with the PHOENIX/BT-Settl \citep{allard2012} stellar atmosphere models. We used informative priors for the effective temperature ($T_{\mathrm{eff}} = 3556 \pm 70$~K, which corresponds to an M2 spectral type), and metallicity ($[\rm{Fe/H}] = -0.06 \pm 0.12$~dex) derived from the co-added HARPS spectra \citep[which we analysed with \specmatch;][]{yee2017}, and for the distance from Gaia. We used non-informative priors for the rest of the parameters. We used an additive jitter \citep{gregory2005} for each set of photometric bands (Gaia, 2MASS, and WISE). The parameters, priors, and posteriors are listed in Table~\ref{table.sed}. The maximum a posteriori (MAP) model is shown in Fig~\ref{fig.sed}. The derived radius ($R_{\star}=0.531^{+0.016}_{-0.012}~\Rsun$) is compatible with the radius computed above using an empirical radius-luminosity relation.

We used the stellar mass and the HARPS stellar rotation period to derive a gyrochronological age, neglecting the influence of the planets. Using the formulation of \citet{barnes2010} and \citet{barneskim2010} with initial periods P$_0$ between 0.12 and 3.4~d, the age is 1.02$^{+0.23}_{-0.15}$~Gyr. We added a 10\% systematic error to the statistical error \citep{meibom2015}. This value agrees with 1.07~$\pm$~0.22~Gyr that was determined based on the rotational period alone using the \mbox{\citet{engle2011}} relation. This was obtained from a sample of M dwarfs.

\begin{table}
\tiny
  \caption{Astrometry, photometry, and stellar parameters for GJ~3090.}\label{table.stellar_parameters}
  \centering
    \setlength{\tabcolsep}{2pt}
\begin{tabular}{lcc}
\hline
\hline
    Parameter & Value & Source \\
     \hline
    Designations & CD-47 399 & \citet{thome1900} \\
                 & HIP 6365 & \citet{turon1993}\\
                 & TIC 262530407 & \citet{stassun2019} \medskip\\
    RA (ICRS, J2000)          & 01$^{\rm h}$ 21$^{\rm m}$ 45.39$^{\rm s}$ & Gaia EDR3 \\
    Dec (ICRS, J2000)         & $-46^{\rm o}$ 42' 51.8''                  & Gaia EDR3 \\
    $\mu$ RA [mas yr$^{-1}$]  & -111.089 $\pm$ 0.020                      & Gaia EDR3 \\
    $\mu$ Dec [mas yr$^{-1}$] & -79.954 $\pm$ 0.025                       & Gaia EDR3 \\
    Parallax [mas]            & 44.555 $\pm$ 0.027                        & Gaia EDR3$^a$ \\
    Distance [pc]             & 22.444 $\pm$ 0.013                        & Sect.~\ref{section.stellar_parameters} \medskip\\

    Gaia-BP      & 11.6698 $\pm$ 0.0032 & Gaia EDR3 \\
    Gaia-G       & 10.5567 $\pm$ 0.0028 & Gaia EDR3 \\
    Gaia-RP      &  9.5053 $\pm$ 0.0029 & Gaia EDR3 \\
    2MASS-J      &  8.168  $\pm$ 0.021  & 2MASS \\
    2MASS-H      &  7.536  $\pm$ 0.036  & 2MASS \\ 
    2MASS-Ks &  7.294  $\pm$ 0.026  & 2MASS \\
    WISE-W1      &  7.053  $\pm$ 0.038  & WISE \\
    WISE-W2      &  7.113  $\pm$ 0.019  & WISE \\
    WISE-W3  &  7.045  $\pm$ 0.016  & WISE \\
    WISE-W4  &  6.941  $\pm$ 0.082  & WISE \medskip\\

    Mass, $M_{\star}$ [\Msun] & 0.519 $\pm$ 0.013 & \citet{mann2019}, Sect.~\ref{sec.photodynamical} \\
    Radius, $R_{\star}$ [\Rsun] & 0.516 $\pm$ 0.016 & \citet{mann2015}, Sect.~\ref{sec.photodynamical} \\
    Effective temperature, $T_{\rm eff}$ [K] & $3556\pm70$ & \specmatch \\
    Metallicity, [Fe/H] [dex] & $-0.06\pm0.12$ & \specmatch \\
    Luminosity [L$_\odot$] & 0.0455$^{+0.0018}_{-0.0016}$ & SED \\
    \vsini{} [\kms] & $\leq 1.468 \pm 0.050$ & $R_{\star}$, P$_{\rm rot}$ \smallskip\\
\hline
\end{tabular}
\tablefoot{$^a$ Corrected according to \citet{lindegren2021}.}
\end{table}

\section{Analysis}\label{section.analysis}

\subsection{Radial velocity}\label{sec.RVanalysis}

\begin{figure}
    \centering  
    \includegraphics[width=0.48\textwidth]{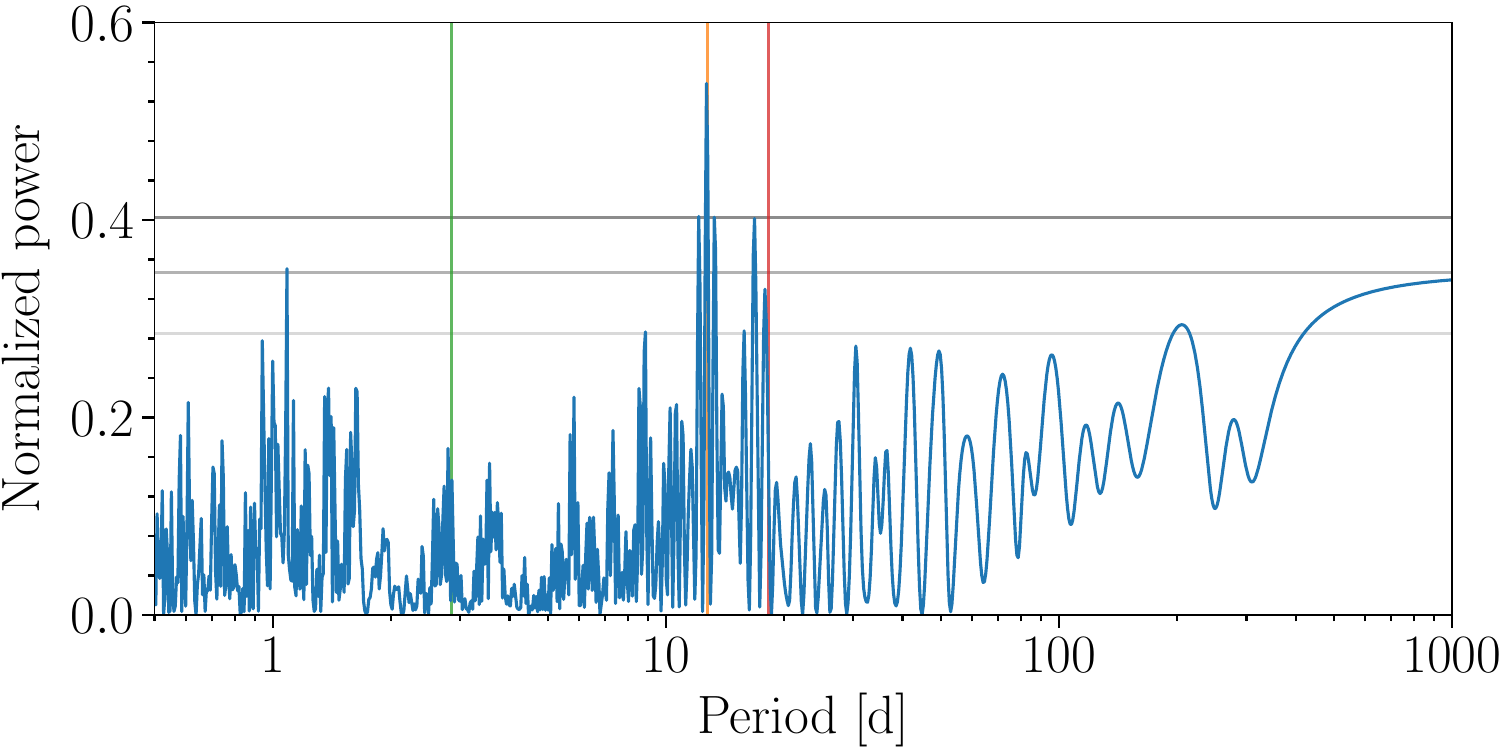} 
    \caption{GLS periodogram of the HARPS velocities for GJ~3090. The horizontal lines represents 10, 1, and 0.1\% false-alarm levels, from bottom to top, respectively. Vertical lines mark the period of the transiting planet (green), the highest peak of the periodogram (orange), and the stellar rotation period estimated in Sect.~\ref{sec.wasp} (red).}
\label{fig.harps}
\end{figure}

\begin{figure}
    \centering  
    \includegraphics[width=0.49\textwidth]{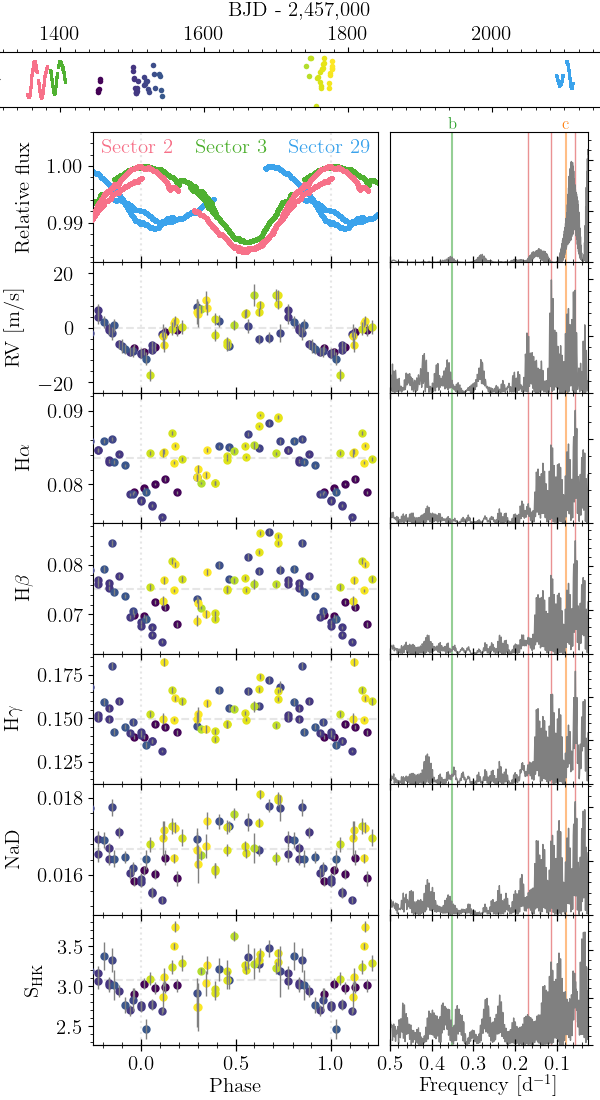} 
    \caption{TESS photometry and HARPS RV and activity indicators folded at the stellar rotational period. To compute the phase, we used the median period of the QPC from the adopted RV modelling (17.733~days) and one arbitrary epoch (BJD~2,458,452.3). The TESS photometry was better folded with a slightly different period at the expense of the RVs. We favoured the general case in which the photometric variability observed by TESS about two years apart could originate from different active regions at different longitudes of the stellar surface. From top to bottom: Times of the TESS (red, green, and blue, depending on the sector) and HARPS observations (at times indicated by the colour of the circles) shifted by the median and scaled with the standard deviation. TESS SAP light curve in which the transits of GJ~3090~b are masked are coloured by sector. HARPS RV. The MAP Keplerians of the adopted RV modelling are subtracted. HARPS H$\alpha$, H$\beta$, H$\gamma$, NaD, and S$_{\rm HK}$ activity indicators (their median is marked by the horizontal line). The panels on the right show the GLS periodogram (in frequency space) of the quantity shown at the left. Vertical lines mark the period of planet~b (green), planet~c (orange), and $P_{\rm rot}$, $P_{\rm rot}/2$, and $P_{\rm rot}/3$ (red). The HARPS observations started 45~days after the end of TESS sector~3 and ended 311~days before the beginning of sector~29.}
\label{fig.activity}
\end{figure}

Figure~\ref{fig.harps} shows the generalised Lomb–Scargle periodogram \citep[GLS,][]{zechmeister2009} of the HARPS RVs. Its highest peak is at 12.7~days, with some power at the stellar rotation period ($P_{\rm rot}$) and at its first harmonic ($P_{\rm rot}/2$). The main peak of the periodogram of the measured activity indicators is at the stellar rotation period, except for S$_{\rm HK}$. For the observations closest in time, the photometric modulation at the stellar rotation period and the RV and activity indicators vary in anti-phase (Fig.~\ref{fig.activity}).

\begin{figure}
    \centering  
    \includegraphics[width=0.49\textwidth]{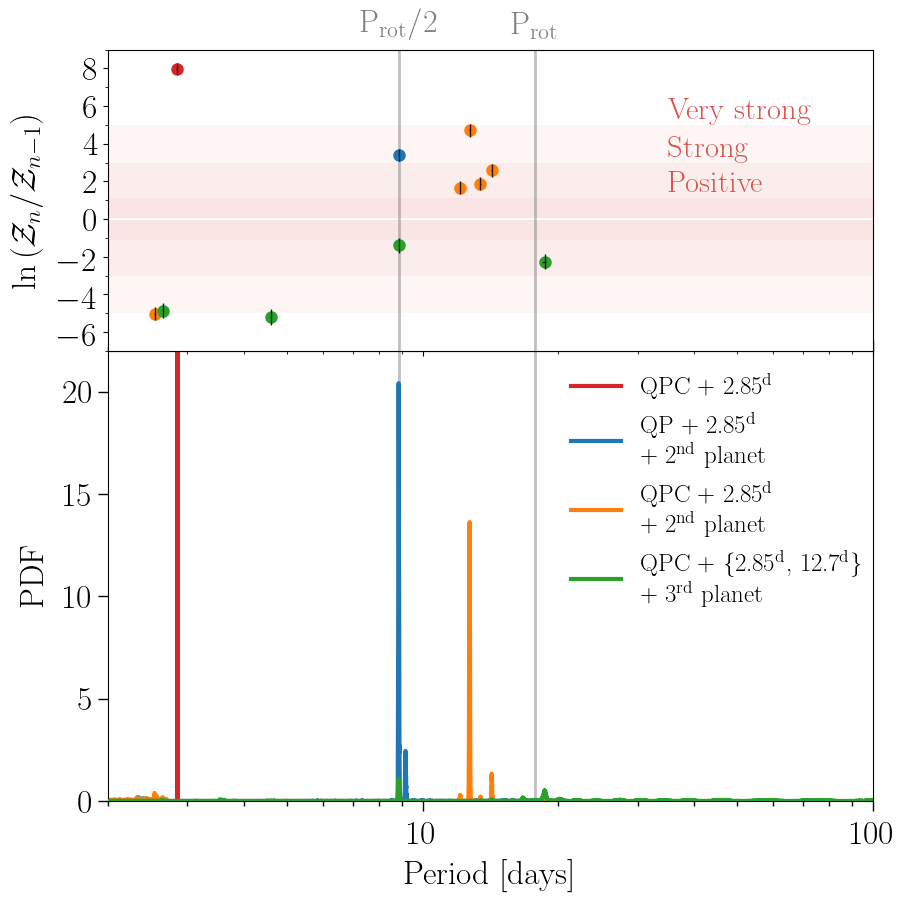} 
    \caption{Search for planets in the HARPS data. Lower panel: Marginal posterior of the trial planet period. Top panel: Bayes factor for the main peaks of the posterior (error bars). The bands are the labelled limits from \citet{kass1995}.}
\label{fig.search}
\end{figure}

To evaluate the significance of the periodicities in the RV series, we iteratively added sine waves describing planets on circular orbits to the model and estimated the evidence. We used \juliet \citep{espinoza2019}, \radvel \citep{fulton2018}, and \dynesty \citep{speagle2020} to model the RV data, and we used a GP regression model to account for the stellar activity signal in the residuals. We first tried the QP kernel GP implemented in \celerite \citep{foreman-mackey2017}, with a prior on the period from Sect.~\ref{sec.wasp}. After adding GJ~3090~b (with a prior on its period and time of conjunction from the transiting TESS candidate), we searched for a second planet within a period range between 3 and 20~days. The posterior orbital period of this second signal peaks at half the stellar rotation period (Fig.~\ref{fig.search}). 

We therefore elected to change the kernel function used in our GP model and instead used the quasi-periodic with cosine (QPC) kernel \citep{perger2021}, which we implemented in \george \citep{ambikasaran2015}. With this new GP model, we searched again for a second planet. As expected, the main peak of the posterior probability for the orbital period now no longer is at half the rotational period of the star, and the distribution presents several peaks around 13~days. The log-Bayes factor ($\mathcal{Z}_n/\mathcal{Z}_{n-1}$, with $\mathcal{Z}_n$ the evidence of the model with $n$ planets, and $\mathcal{Z}_{n-1}$ the evidence of the model with $n-1$ planets) for this model with respect to the one-planet model is 5.79~$\pm$~0.23, which can be interpreted as very strong evidence \citep{kass1995} in favour of this model (as we assume equal probability for the $n$ and $n-1$ models, the Bayes factor is equivalent to the odds ratio).

The marginal posterior distribution for the orbital period of the second planet presents a number of peaks. The largest peak lies at 12.7 days, and the remaining peaks are compatible with aliases (see Fig.~\ref{fig.alias}). We nevertheless estimated the contribution of each peak to the total evidence value. The peak at 12.7 days holds the largest evidence, and the remaining peaks have log-Bayes factors with respect to the main peak of -3.06~$\pm$~0.49 for the 12.1~days, -2.83~$\pm$~0.49 for the 13.4~days and -2.12~$\pm$~0.49 for the 14.2~days. We adopted the 12.7-day period for the second planet.

We searched for a third planet within a period range between 2 (at less than two~days, multiple aliases appear at and around 1.0 and 1.5~days, see Fig.~\ref{fig.harps}) and 2.7~days, 3 to 12~days, and 13.1 to 100~days and found log-Bayes factors with respect to the two-planet model of -2.39~$\pm$~0.28, -0.71~$\pm$~0.28, and -0.33~$\pm$~0.28, respectively. We conclude that the current data do not favour a model with a third planet. The marginal posterior distribution for the orbital period for the third planet peak again at half the stellar rotation period despite the use of the QPC kernel.   

If instead of the described scheme we start by modelling the 12.7-day signal and then search for a second signal, the marginal posterior distribution of the trial period in the 2 to 12~days range peaks at 2.881~days (an alias of the 2.853~days period of GJ~3090~b, which is also present in the periodogram). We also tested that the model with the QPC kernel and a planet with the priors from the transiting TESS candidate is favoured, $\ln(\mathcal{Z}_{\rm QPC + 2.85^d}/\mathcal{Z}_{\rm QPC}) = 7.43 \pm 0.31$, against a model with only the QPC kernel (Fig.~\ref{fig.search}). 

\begin{figure}
    \centering  
    \includegraphics[width=0.48\textwidth]{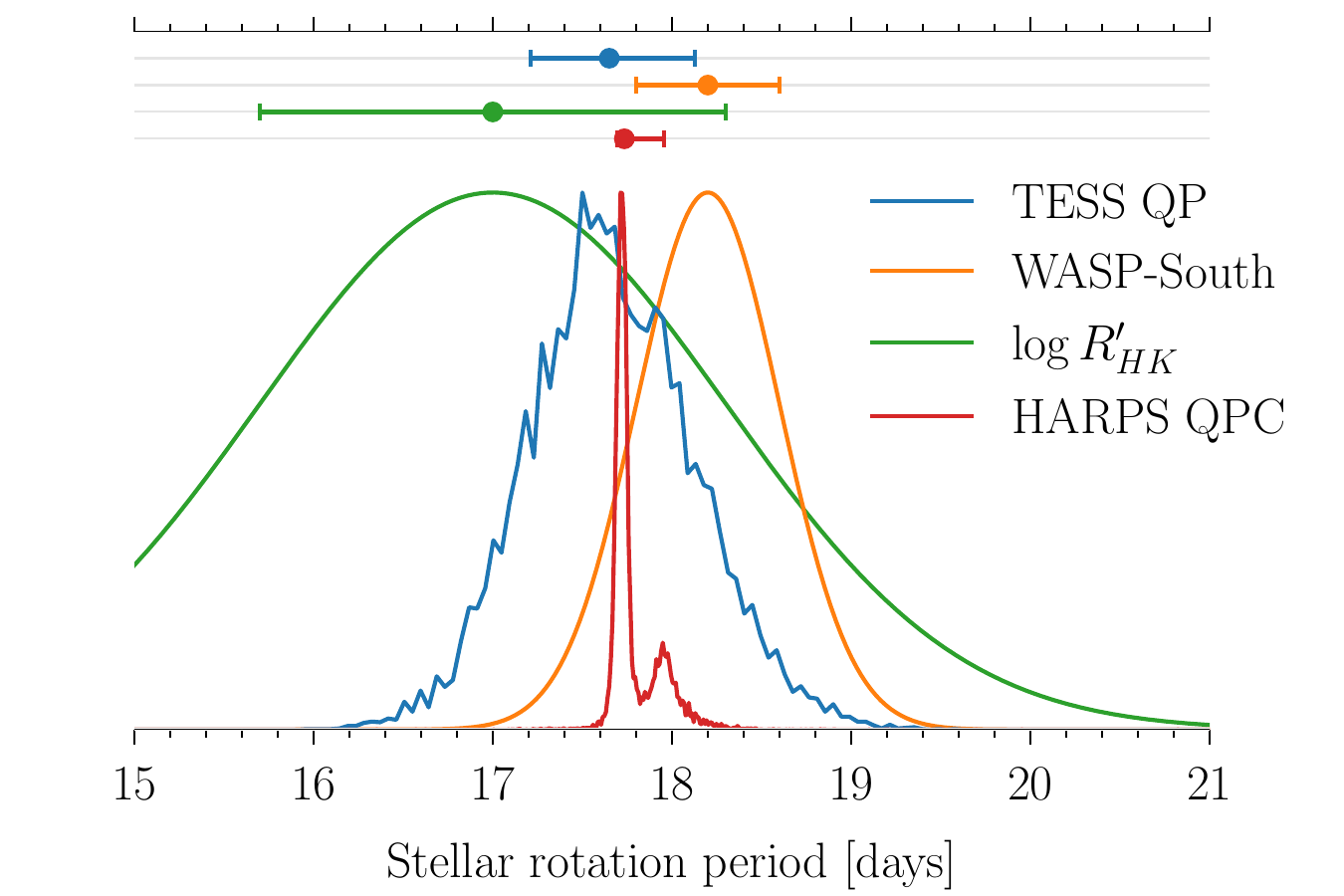} 
    \caption{Estimates of the stellar rotation period. The distributions are normalised by their maximum for visibility purposes. Error bars at the top of the distributions represent the median and the 68.3\% CI. The distributions are assumed normal for the periods derived from WASP-South and $\log R'_{HK}$.}
\label{fig.Prot}
\end{figure}

In order to study the influence of the eccentricity on the RV semi-amplitudes, we finally modelled the RV with two planets on eccentric orbits. We used non-informative priors for all the parameters except for the period and time of conjunction of GJ~3090~b. The parameter priors, posterior medians, and 68.3\% credible intervals (CIs) are shown in Table~\ref{table.RV}. The period of the QPC kernel parameter is well constrained although it has an uniform prior in this analysis. This is even the most precise of all the fully compatible stellar rotation period estimates presented in Sect.~\ref{section.observations} (Fig.~\ref{fig.Prot}). The parameter $h_2$ of the QPC kernel is poorly constrained, even though it is key for the $P_{\rm rot}/2$ signal.

\begin{figure}
    \centering  
    \includegraphics[width=0.48\textwidth]{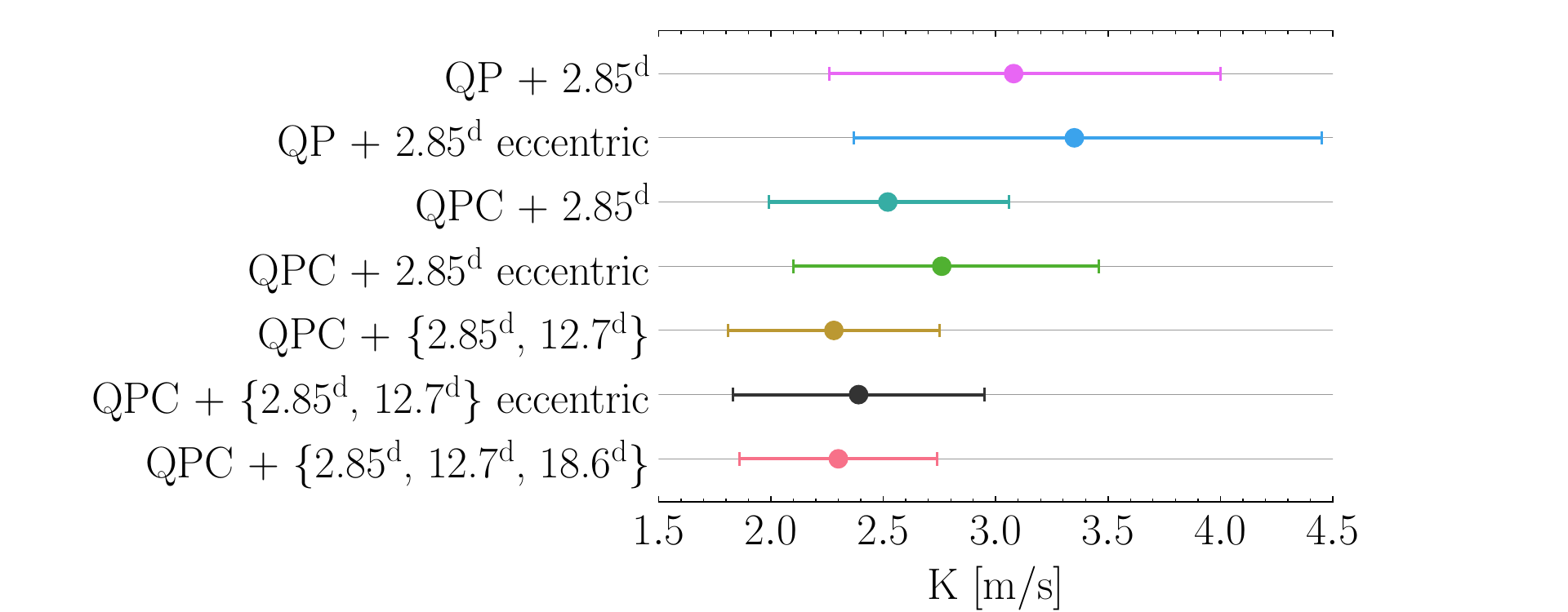} 
    \caption{RV semi-amplitude due to GJ~3090~b from different models. Our preferred model is shown in black.}
\label{fig.K}
\end{figure}

We investigated how the RV semi-amplitude of GJ~3090~b changed with the model assumption. We summarise the results in Fig.~\ref{fig.K}. The RV semi-amplitude and its uncertainty can change by up to 40\% and 86\% (relative change when the adopted value is taken as reference), respectively. 

\subsection{Keplerian modelling}\label{sec.juliet}

We first adjusted a Keplerian model to the TESS, Spitzer, ExTrA, and LCOGT transit photometry and the HARPS RVs using \juliet \citep{espinoza2019}. \juliet uses \batman \citep{kreidberg2015} for the transit model and \radvel \citep{fulton2018} to model the RVs. For the transit photometry, a GP was used to model the residuals, with the approximate Matern or QP kernels (depending on the data set) included in \texttt{celerite} \citep{foreman-mackey2017}. We oversampled the TESS and LCOGT models by a factor 5 \citep{kipping2010} because the averaging times of both data sets are longer than one~minute. For the RV, we used the QPC kernel \citep{perger2021} with \george \citep{ambikasaran2015}. We used a normal prior for the stellar density (5.29~$\pm$~0.53~$\mathrm{g\;cm^{-3}}$) from the stellar mass and radius derived in Sect.~\ref{section.stellar_parameters} and non-informative priors for the remaining parameters. We sampled from the posterior with \dynesty \citep{speagle2020}. In Table~\ref{table.results} we list the prior, the median, and the 68\% CI of the marginal distributions of the inferred system parameters. We used the stellar mass and radius derived in Sect.~\ref{section.stellar_parameters} to compute the planetary masses and the radius of planet~b. Figure~\ref{fig.juliet} shows the data sets and the model from this analysis. 

\begin{table*}
  \scriptsize
\renewcommand{\arraystretch}{1.25}
\centering
\caption{Inferred system parameters.}\label{table.results}
\begin{tabular}{lcccccc}
\hline
\hline
 & & Keplerian & Keplerian & Dynamical & Dynamical & Stable \\
Parameter & Units & Prior &  Median and 68.3\% CI & Prior & Median and 68.3\% CI & Median and 68.3\% CI\\
\hline
\emph{Star} \\
Mass, $M_\star$              & [\Msun]                 &                 &                 & $N(0.519, 0.013)$      & $0.520 \pm 0.013$      & $0.519 \pm 0.013$   \\
Radius, $R_\star$            & [\Rnom]                 &                 &                 & $N(0.517, 0.016)$      & $0.517 \pm 0.016$      & $0.516 \pm 0.016$    \\
Mean density, $\rho_{\star}$ & [$\mathrm{g\;cm^{-3}}$] & $N(5.29, 0.53)$ & $5.36 \pm 0.53$ &                        & $5.30^{+0.52}_{-0.45}$ & $5.32^{+0.52}_{-0.47}$   \\
Surface gravity, \logg\      & [cgs]                   &                 &                 &                        & $4.727 \pm 0.028$      & $4.727 \pm 0.029$         \\
\citet{kipping2013} $q_1$ TESS &                       & $U(0, 1)$       & $0.21^{+0.23}_{-0.12}$ & $U(0, 1)$       & $0.33^{+0.29}_{-0.18}$ & $0.34^{+0.28}_{-0.18}$ \\
\citet{kipping2013} $q_2$ TESS &                       & $U(0, 1)$       & $0.22^{+0.39}_{-0.17}$ & $U(0, 1)$       & $0.18^{+0.31}_{-0.14}$ & $0.18^{+0.30}_{-0.13}$ \smallskip\\

\emph{Planet~b} \\
Semi-major axis, $a$                   & [au]             &            & $0.0317 \pm 0.0014$ &                 & $0.03165 \pm 0.00027$ & $0.03165 \pm 0.00027$ \\
Eccentricity, $e$                      &                  &            & $0.18 \pm 0.12$ &             & $0.16 \pm 0.11$       & < 0.32$^\dagger$, ($0.15 \pm 0.11$) \\
Argument of pericentre, $\omega$       & [\degree]        &            & $146^{+68}_{-53}$ &                 & $142^{+74}_{-57}$ & $137^{+73}_{-57}$ \\
Inclination, $i$                       & [\degree]        &            & $87.09^{+0.98}_{-0.29}$ & $S(0, 90)$      & $87.13^{+0.80}_{-0.29}$ & $87.14^{+0.79}_{-0.30}$ \\
Longitude of the ascending node, $\Omega$ & [\degree]     &            & & & 180 (fixed at $t_{\mathrm{ref}}$) & 180 (fixed at $t_{\mathrm{ref}}$) \\
Mean anomaly, $M_0$                    & [\degree]        &            & &                 & $327^{+40}_{-73}$ & $331^{+41}_{-69}$ \\
$\sqrt{e}\cos{\omega}$                 &                  & $U(-1, 1)$ & $-0.25^{+0.26}_{-0.17}$ & $U(-1, 1)$      & $-0.21^{+0.26}_{-0.18}$ & $-0.19^{+0.26}_{-0.19}$ \\
$\sqrt{e}\sin{\omega}$                 &                  & $U(-1, 1)$ & $0.14 \pm 0.32$ & $U(-1, 1)$      & $0.17^{+0.27}_{-0.34}$ & $0.17^{+0.27}_{-0.31}$ \\
Mass ratio, $M_{\mathrm{p}}/M_\star$   &                  &            & & $U(0, 1)$       & $(19.3 \pm 4.2)\times 10^{-6}$ & $(19.4 \pm 4.1)\times 10^{-6}$ \\
Radius ratio, $R_{\mathrm{p}}/R_\star$ &                  &            & $0.0384^{+0.0011}_{-0.0015}$ & $U(0, 1)$       & $0.0379^{+0.0012}_{-0.0014}$ & $0.0379^{+0.0011}_{-0.0014}$ \\
Scaled semi-major axis, $a/R_{\star}$  &                  &            & $13.22 \pm 0.45$ &                 & $13.16 \pm 0.42$ & $13.18 \pm 0.42$\\
Impact parameter, $b$                  &                  &            & $0.650^{+0.093}_{-0.32}$ &                 & $0.633^{+0.099}_{-0.27}$ & $0.631^{+0.093}_{-0.26}$ \\
Transit duration, $T_{14}$             & [h]              &            & $1.270 \pm 0.022$  & & $1.281 \pm 0.025$   & $1.281 \pm 0.024$ \\
\citet{espinoza2018} $r_1$             &                  & $U(0, 1)$  & $0.766^{+0.062}_{-0.21}$ &                 & & \\
\citet{espinoza2018} $r_2$             &                  & $U(0, 1)$  & $0.0384^{+0.0011}_{-0.0015}$ &                 & & \\
T$_0$ ($T_0'$)\;-\;2\;450\;000 & [BJD$_{\mathrm{TDB}}$]   & $U(8370.417, 8370.420)$ & $8370.41849 \pm 0.00030$ & $U(7370, 9370)$ & $8370.41847 \pm 0.00034$ & $8370.41849 \pm 0.00034$ \\
Orbital period, P ($P'$)               & [d]              & $U(2.85309, 2.85312)$ & $2.8531054 \pm 0.0000023$ & $U(0, 1000)$    & $2.853134^{+0.000066}_{-0.000047}$ & $2.853136^{+0.000064}_{-0.000038}$ \\
RV semi-amplitude, K ($K'$)            & [\ms]            & $U(0, 8)$  & $2.36 \pm 0.52$ &                 & $2.37 \pm 0.51$ & $2.38 \pm 0.51$ \\

Mass, $M_{\mathrm{p}}$                 &[\MEarth]         &            & $3.31 \pm 0.73$ &                 & $3.33 \pm 0.72$ & $3.34 \pm 0.72$ \\
Radius, $R_{\mathrm{p}}$               &[\Renom]          &            & $2.16 \pm 0.10$ &                 & $2.13 \pm 0.11$ & $2.13 \pm 0.11$ \\
Mean density, $\rho_{\mathrm{p}}$ &[$\mathrm{g\;cm^{-3}}$]&            & $1.80^{+0.50}_{-0.42}$ &                 & $1.87^{+0.54}_{-0.44}$ & $1.89^{+0.52}_{-0.45}$ \\
Surface gravity, $\log$\,$g_{\mathrm{p}}$ &[cgs]          &            & $2.841^{+0.095}_{-0.11}$ &                 & $2.853^{+0.098}_{-0.11}$ & $2.857^{+0.096}_{-0.11}$ \\
Equilibrium temperature, T$_{\rm eq}$  & [K]              &            & $692 \pm 18$ &                 & $693 \pm 17$ & $693 \pm 18$ \smallskip\\

\emph{Planet~c} \\
Semi-major axis, $a$                   & [au]             &            & $0.0860 \pm 0.0039$ &                 & $0.08578 \pm 0.00075$ & $0.08575 \pm 0.00074$ \\
Eccentricity, $e$                      &                  &            & $0.23^{+0.12}_{-0.13}$ &                 & $0.17 \pm 0.11$ & < 0.31$^\dagger$, ($0.16 \pm 0.11$) \\
Argument of pericentre, $\omega$       & [\degree]        &            & $357^{+0.31}_{-0.37}$ &                 & $0^{+44}_{-53}$ & $-1^{+45}_{-57}$ \\
Inclination, $i$                       & [\degree]        &            & & $S(0, 180)$     & $85 \pm 55$ & $86 \pm 46$ \\
Longitude of the ascending node, $\Omega$ & [\degree]     &            & & $U(90, 270)$    & $157^{+70}_{-44}$ & $157^{+50}_{-39}$ \\
Mean anomaly, $M_0$                    & [\degree]        &            & &                 & $68 \pm 47$ & $68 \pm 51$ \\
$\sqrt{e}\cos{\omega}$                 &                  & $U(-1, 1)$ & $0.42^{+0.14}_{-0.26}$ & $U(-1, 1)$      & $0.33^{+0.15}_{-0.32}$ & $0.32^{+0.15}_{-0.33}$ \\
$\sqrt{e}\sin{\omega}$                 &                  & $U(-1, 1)$ & $-0.02 \pm 0.22$ & $U(-1, 1)$      & $0.00 \pm 0.23$ & $0.00 \pm 0.22$ \\
Mass ratio, $M_{\mathrm{p}}/M_\star$   &                  &            & & $U(0, 1)$       & $(10.8^{+8.5}_{-2.5})\times 10^{-5}$ & $(9.9^{+5.1}_{-1.9})\times 10^{-5}$ \\
Scaled semi-major axis, $a/R_{\star}$  &                  &            & $35.8 \pm 1.2$    &                 & $35.7 \pm 1.1$ & $35.7 \pm 1.1$ \\
Impact parameter, $b$                  &                  &            & &                 & $22.2^{+9.4}_{-14}$ & $19^{+10}_{-13}$ \\
T$_0$ ($T_0'$)\;-\;2\;450\;000  & [BJD$_{\mathrm{TDB}}$]  & $U(8367, 8376)$ & $8370.39^{+1.3}_{-0.90}$ & $U(7370, 9370)$ & $8370.92^{+1.1}_{-0.90}$ & $8370.96^{+1.2}_{-0.90}$ \\
Orbital period, P ($P'$)               & [d]              & $U(12.5, 12.9)$ & $12.736^{+0.023}_{-0.030}$ & $U(0, 1000)$    & $12.730^{+0.025}_{-0.031}$ & $12.729^{+0.025}_{-0.031}$ \\
RV semi-amplitude, K ($K'$)            & [\ms]            & $U(0, 20)$ & $6.23 \pm 0.75$ &                 & $6.05 \pm 0.69$ & $6.01 \pm 0.69$ \\

Mass, $M_{\mathrm{p}}$                 &[\MEarth]         &            & &                 & $18.6^{+15}_{-4.4}$ & $17.1^{+8.9}_{-3.2}$ \\
$M_{\mathrm{p}}\sin{i}$                &[\MEarth]         &            & $14.3 \pm 1.5$ &                 & $14.0 \pm 1.6$ & $13.9 \pm 1.6$ \\
Equilibrium temperature, T$_{\rm eq}$  & [K]              &            & $420 \pm 11$ &                 & $421 \pm 11$ & $421 \pm 11$ \smallskip\\

Mutual inclination, $i_\mathrm{rel}$   & [\degree]        &            & &                 & $66^{+16}_{-27}$ & $59^{+16}_{-26}$ \smallskip\\

\hline
\end{tabular}
\tablefoot{The table lists the prior, posterior median, and 68.3\% CI for the parameters of the Keplerian model (Sect.~\ref{sec.juliet}) and of the dynamical model, without and with the stability constraint (Sect.~\ref{sec.photodynamical}). For the dynamical model, the Jacobi orbital elements are given for the reference time $t_{\mathrm{ref}}~=~2\;458\;370.418461$~BJD$_{\mathrm{TDB}}$. $^\dagger$ Upper limit, 95\% confidence. The planetary equilibrium temperature is computed with $T_{\mathrm{eff}} = 3556 \pm 70$~K and for zero albedo and full day-night heat redistribution. T$_0$ is the time of conjunction. $T'_0 \equiv t_{\mathrm{ref}} - \frac{P'}{2\pi}\left(M_0-E+e\sin{E}\right)$ with $E=2\arctan{\left\{\sqrt{\frac{1-e}{1+e}}\tan{\left[\frac{1}{2}\left(\frac{\pi}{2}-\omega\right)\right]}\right\}}$, $P' \equiv \sqrt{\frac{4\pi^2a^{3}}{\mathcal G M_{\star}}}$, $K' \equiv \frac{M_p \sin{i}}{M_\star^{2/3}\sqrt{1-e^2}}\left(\frac{2 \pi \mathcal G}{P'}\right)^{1/3}$. CODATA 2018: $\mathcal G$ = 6.674$\;$30\ten[-11]~$\rm{m^3\;kg^{-1}\;s^{-2}}$. IAU 2012: \rm{au} = 149$\;$597$\;$870$\;$700~\rm{m}$\;$. IAU 2015: \Rnom = 6.957\ten[8]~\rm{m}, \GMnom = 1.327$\;$124$\;$4\ten[20]~$\rm{m^3\;s^{-2}}$, \Renom~=~6.378$\;$1\ten[6]~\rm{m}, \GMenom = 3.986$\;$004\ten[14]~$\rm{m^3\;s^{-2}}$, $\Msun$ = \GMnom/$\mathcal G$, \MEarth = \GMenom/$\mathcal G$, $k^2$ = \GMnom$\;(86\;400~\rm{s})^2$/$\rm{au}^3$. $N(\mu, \sigma)$: Normal distribution with mean $\mu$, and standard deviation $\sigma$. $U(a, b)$: A uniform distribution defined between a lower $a$ and upper $b$ limit. $S(a, b)$: A Sinusoidal distribution defined between a lower $a$ and upper $b$ limit. $J(a, b)$: Jeffreys (or log-uniform) distribution defined between a lower $a$ and upper $b$ limit.}
\end{table*}

\setcounter{table}{2}
\begin{table*}
  \scriptsize
\renewcommand{\arraystretch}{1.25}
\centering
\caption{continued.}
\begin{tabular}{lcccccc}
\hline
\hline
 & & Keplerian & Keplerian & Dynamical & Dynamical & Stable \\
Parameter & Units & Prior &  Median and 68.3\% CI & Prior & Median and 68.3\% CI & Median and 68.3\% CI\\
\hline

\emph{HARPS} \\
Systemic velocity, $\gamma$  &[\kms] & $U(17.385, 17.435)$ & $17.4095 \pm 0.0046$ & $U(-100, 100)$ & $17.4095 \pm 0.0043$ & $17.4095 \pm 0.0041$ \\
Jitter                &[\ms]    & $J(0.01, 4)$ & $0.101^{+0.41}_{-0.080}$ & $U(0, 1000)$   & $0.66^{+0.67}_{-0.45}$ & $0.68^{+0.65}_{-0.46}$ \\
QPC $h_1$             & [\ms]   & $J(0.01, 30)$ & $9.5^{+3.1}_{-2.2}$ & $J(0.01, 30)$  & $9.1^{+3.2}_{-2.3}$ & $9.0^{+3.0}_{-2.2}$ \\
QPC $h_2$             & [\ms]   & $J(0.01, 30)$ & $0.52^{+3.8}_{-0.49}$ & $J(0.01, 30)$  & $0.55^{+3.6}_{-0.51}$ & $0.59^{+3.7}_{-0.55}$ \\
QPC $P$               & [d]     & $U(17, 19)$  & $17.729^{+0.17}_{-0.036}$ & $U(17, 19)$    & $17.732^{+0.17}_{-0.038}$ & $17.731^{+0.15}_{-0.038}$ \\
QPC $\lambda$         & [d]     & $U(1, 2000)$ & $640^{+410}_{-350}$ & $U(1, 2000)$   & $600^{+410}_{-320}$ & $610^{+390}_{-320}$ \smallskip\\

\emph{TESS} \\
Relative flux        & [Relative flux] &            & & $U(0.9, 1.1)$     & $1.000003 \pm 0.000023$ & $1.000002 \pm 0.000022$\\
Jitter               & [ppm]           &            & & $J(1, 10000)$     & $9.0^{+33}_{-7.0}$ & $9.4^{+33}_{-7.3}$ \\
Timescale of the GP & [days]          &            & & $J(0.001, 1000)$  & $1.9^{+190}_{-1.9}$ & $1.8^{+180}_{-1.7}$ \\
Amplitude of the GP  & [ppm]           &            & & $J(1, 10^{6})$    & $6.9^{+23}_{-5.1}$ & $6.6^{+22}_{-4.8}$ \\
\hline
\end{tabular}
\end{table*}

In the next section we use a self-consistent Newtonian model that takes the gravitational interactions between the planets into account. Even though we detect no interactions, accounting for them allows us to exclude combinations of system parameters that would result in detectable interactions and constrain the true mass of planet~c.

\subsection{Dynamical modelling}\label{sec.photodynamical}

We modelled the TESS photometry and HARPS RV measurements and accounted for the gravitational interactions between the adopted components of the system using a photodynamical model. The TESS transits are the most numerous of the transit observations and dominate the determination of the planet-to-star radius ratio. For simplicity, we did not model the transit observations with other instruments. The planet positions and velocities in time were obtained through an n-body integration. The sky-projected positions were used to compute the light curve \citep{mandelagol2002} using a quadratic limb-darkening law \citep{manduca1977} that we parametrised following \citet{kipping2013}. The light-time effect \citep{irwin1952} was included in the model. To account for the integration time, the model was oversampled by a factor 3 and was then binned back to match the cadence of the data points \citep{kipping2010}. The line-of-sight projected velocity of the star issued from the n-body integration was used to model the RV measurements. We used the n-body code \rebound \citep{rein2012} with the \whf integrator \citep{rein2015} and an integration step of 0.005~days, which resulted in a model error smaller than 1~ppm for the photometric model. The model was parametrised using the stellar mass and radius, planet-to-star mass ratios, planet~b-to-star radius ratio, and Jacobi orbital elements (Table~\ref{table.results}) at reference time, $t_{\mathrm{ref}}~=~2\;458\;370.418461$~BJD$_{\mathrm{TDB}}$. Due to the symmetry of the problem, we fixed the longitude of the ascending node of the interior planet $\Omega_{\mathrm{b}}$ at $t_{\mathrm{ref}}$ to 180\degree, and we limited its inclination to $i_c<90$\degree. 
We used a GP regression model with an approximate Matern kernel \citep[\celerite,][]{foreman-mackey2017} for the model of the error terms of the transit light curves, and the QPC kernel \citep{perger2021, ambikasaran2015} for the RVs.
For the photometric dataset, we added a transit normalisation factor and a jitter parameter. For the RV, we added a systemic RV and a jitter parameter. In total, the model has 28 free parameters. We used normal priors for the stellar mass and radius from Sect.~\ref{section.stellar_parameters}, non-informative sinusoidal priors for the orbital inclinations (uniform in $\cos{i}$), and non-informative uniform prior distributions for the rest of the parameters. The joint posterior distribution was sampled using the \emcee\ algorithm \citep{goodmanweare2010, emcee} with 200~walkers with starting points based on the results of Sect.~\ref{sec.juliet}. We ran 1.2$\e{6}$ steps of the \emcee\ algorithm and used the last 200\;000 steps for the final inference. In Table~\ref{table.results} we list the prior, the median, and the 68\% CI of the inferred marginal distributions of the system parameters.

\begin{figure}
  \centering
  \includegraphics[width=0.49\textwidth]{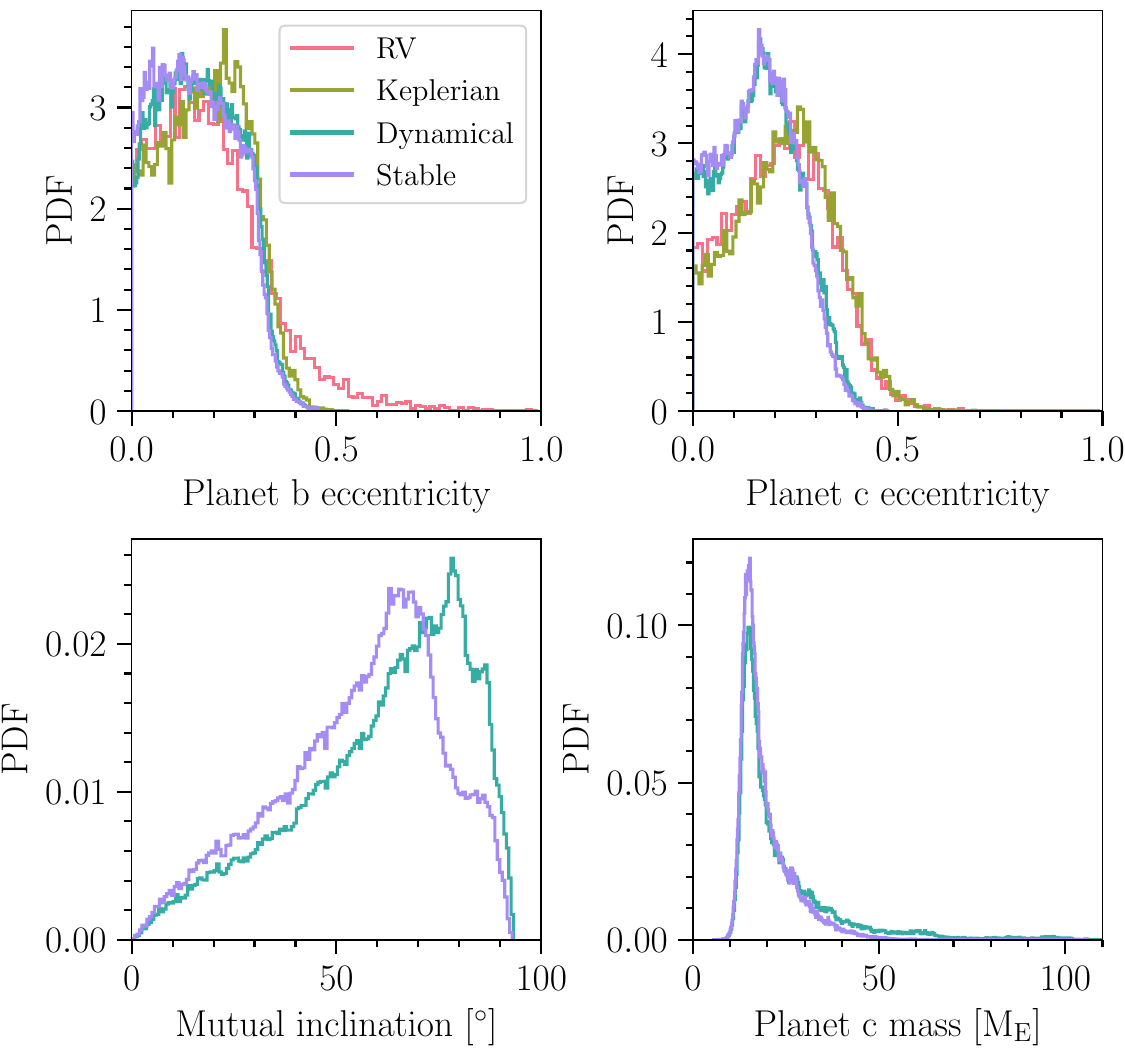}
  \caption{Posterior distribution comparison for the planet eccentricities, the mutual inclination between the planet orbits, and the mass of planet~c before (dynamical) and after (stable) the stability constraint. For the eccentricities, the posteriors for the RV analysis (Sect.~\ref{sec.RVanalysis}) and Keplerian modelling (Sect.~\ref{sec.juliet}) are also shown.} \label{fig.stability}
\end{figure}

We used the mean exponential growth of nearby orbits \citep[MEGNO;][]{cincotta2000, cincotta2003}, a chaos indicator that is implemented in \rebound, to assess the stability of the system. The period ratio of the planets is 4.461~$\pm$~0.011, close to the 9:2 seventh-order mean motion resonance. For each sample of the posterior, we ran \rebound with the \whf integrator for 10$^4$ orbits of planet~c with time steps of 0.1~days and computed the MEGNO value. We followed \citet{hadden2018} and considered stable the solutions with a MEGNO value between 0 and 2.3 (\citealt{hadden2018} ran the simulation for 3000 orbits of the outer planet, with period $P'$, and considered regular trajectories, $\frac{3000 P'}{{\rm MEGNO}} > 1300 P'$, which corresponds to $0 < {\rm MEGNO} \lesssim 2.3$). This represents 77\% of the posterior samples. The requirement of stability removes some extreme values of mutual inclination and planet~c mass from the posterior (Fig.~\ref{fig.stability}). We added a column to Table~\ref{table.results} based on the stable samples of the posterior, which we adopt as our preferred values. We were only able to determine upper limits for the eccentricities of planet~b and c of 0.32 and 0.31, respectively, at 95\% CI. The mutual inclination between the planets b and c is poorly constrained by the observations.

\begin{figure*}
  \centering
  \includegraphics[width=0.88\textwidth]{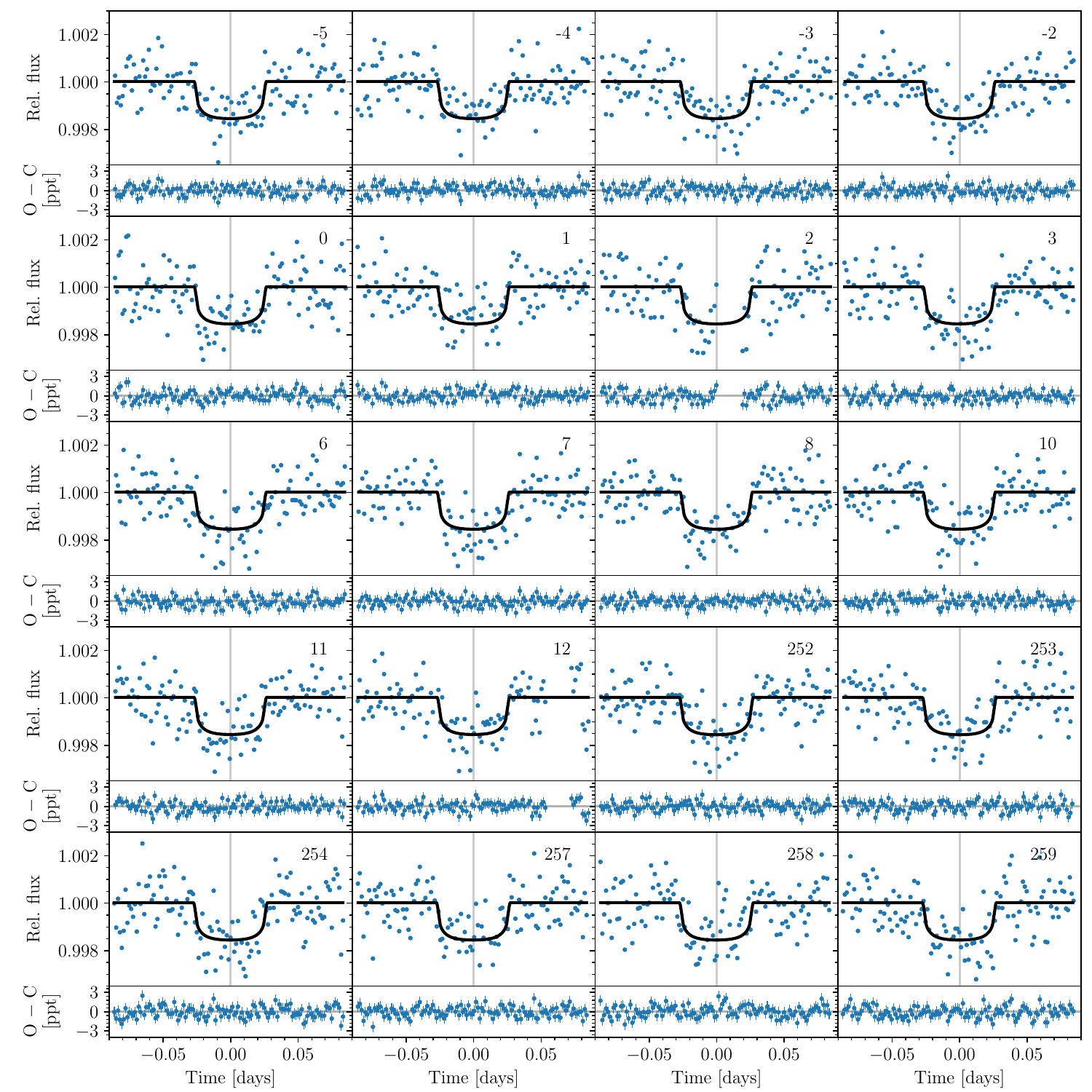}
  \caption{Transits of GJ~3090~b observed with TESS (blue points) and the MAP model (black line). Each panel is centred at the linear ephemeris from Sect.~\ref{sec.juliet} (indicated by the vertical grey lines). Each panel is labelled with the epoch; zero is the transit closest to $t_{\mathrm{ref}}$. In the lower part of each panel, the residuals after subtracting the MAP model are shown.} \label{fig.PH}
\end{figure*}

\begin{figure*}
  \centering
  \includegraphics[width=0.9\textwidth]{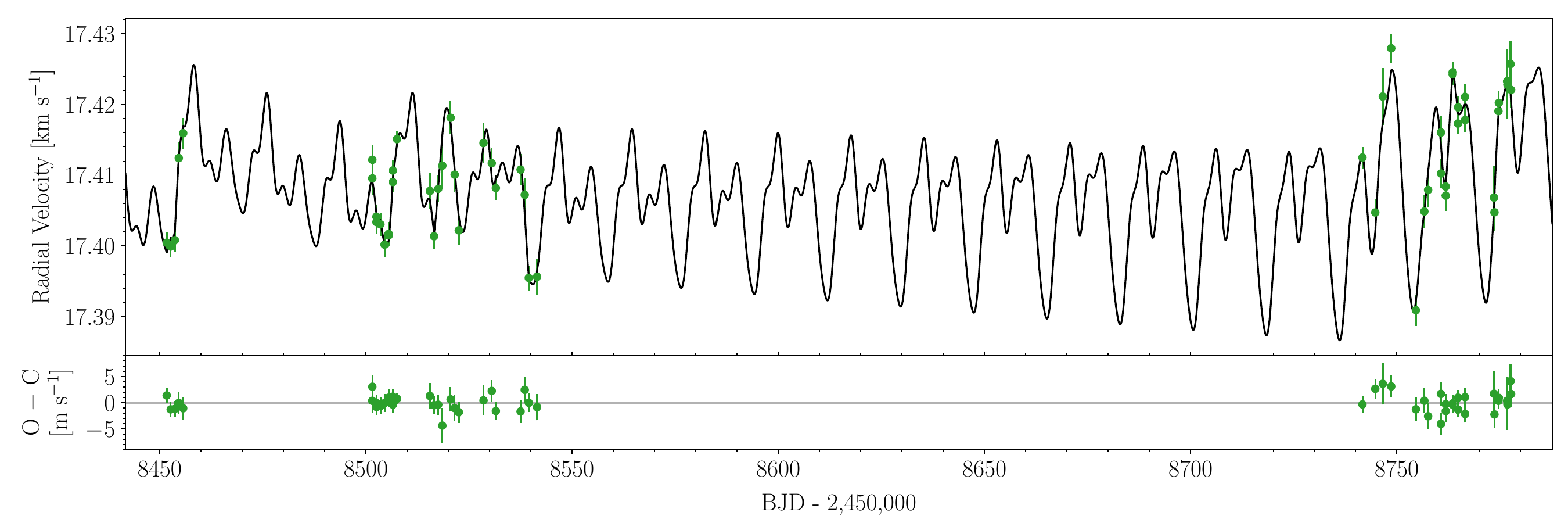}
  \caption{HARPS RVs of GJ~3090 (green error bars) and the MAP model (black line). In the lower panel, the residuals after subtracting the MAP model are shown.} \label{fig.RV}
\end{figure*}

The one- and two-dimensional projections of the posterior sample are shown in Fig.~\ref{fig.pyramid}. The MAP model is plotted in Figs.~\ref{fig.PH} and \ref{fig.RV}.
The posteriors of the transit-timing variations (TTVs) are shown in Fig.~\ref{fig.TTVs}, along with individually determined TESS\ transit times using \juliet. 


\section{Results and discussion}\label{section.results}

\begin{figure}
  \centering
  \includegraphics[width=0.49\textwidth]{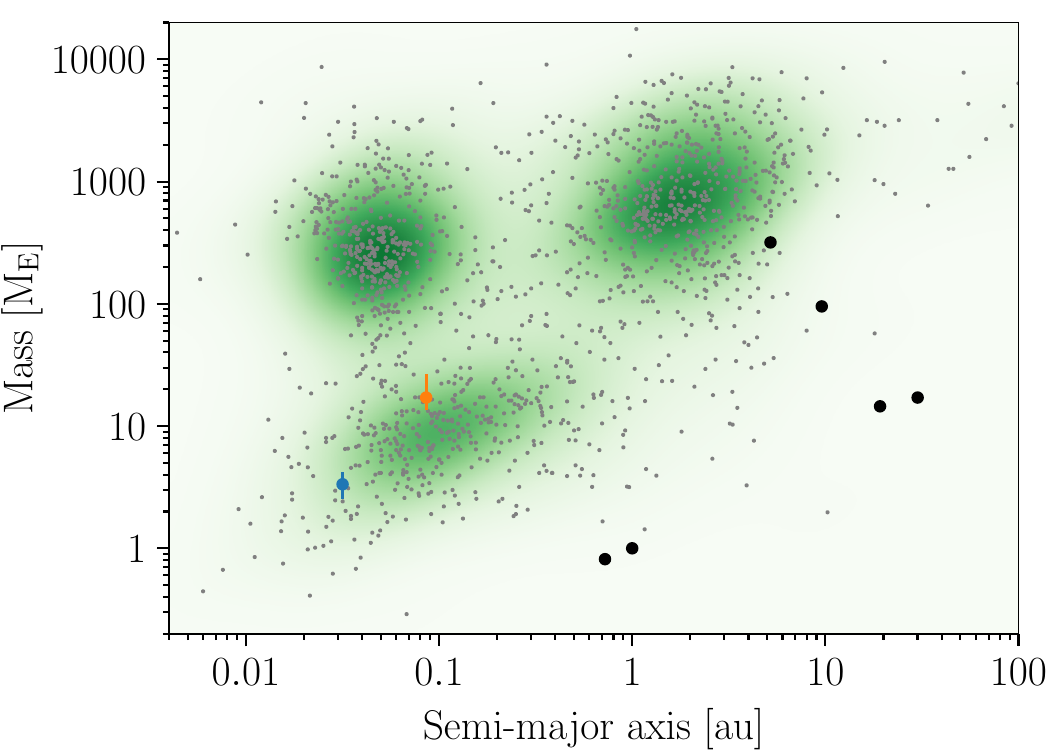}
  \caption{Planet mass vs orbital semi-major axis in logarithmic scale. Error bars mark the position of planet~b (blue) and planet~c (orange). Grey dots are planets listed in the NASA Exoplanet Archive accessed through DACE API (\protect\url{dace.unige.ch}). A Gaussian kernel density estimate is shown in different intensities of green. Solar System planets (black points, from left to right: Venus, Earth, Jupiter, Saturn, Uranus, and Neptune) are taken from NASA.} \label{fig.MassSeparation}
\end{figure}

The GJ~3090 system is composed of an M2-dwarf star orbited by a transiting mini-Neptune, GJ~3090~b, with a period of 2.9~days (Fig.~\ref{fig.MassSeparation}). We found evidence for an outer Neptune-mass planet candidate in the RV data whose transits are not detected in TESS data (Fig.~\ref{fig.tess}). Below we investigate the transiting planet GJ~3090~b further.

\subsection{Characterisation of the interior}\label{sec.interior}

\begin{figure}[h]
\centering
  \begin{tabular}{@{}cc@{}}
    \includegraphics[width=0.48\textwidth]{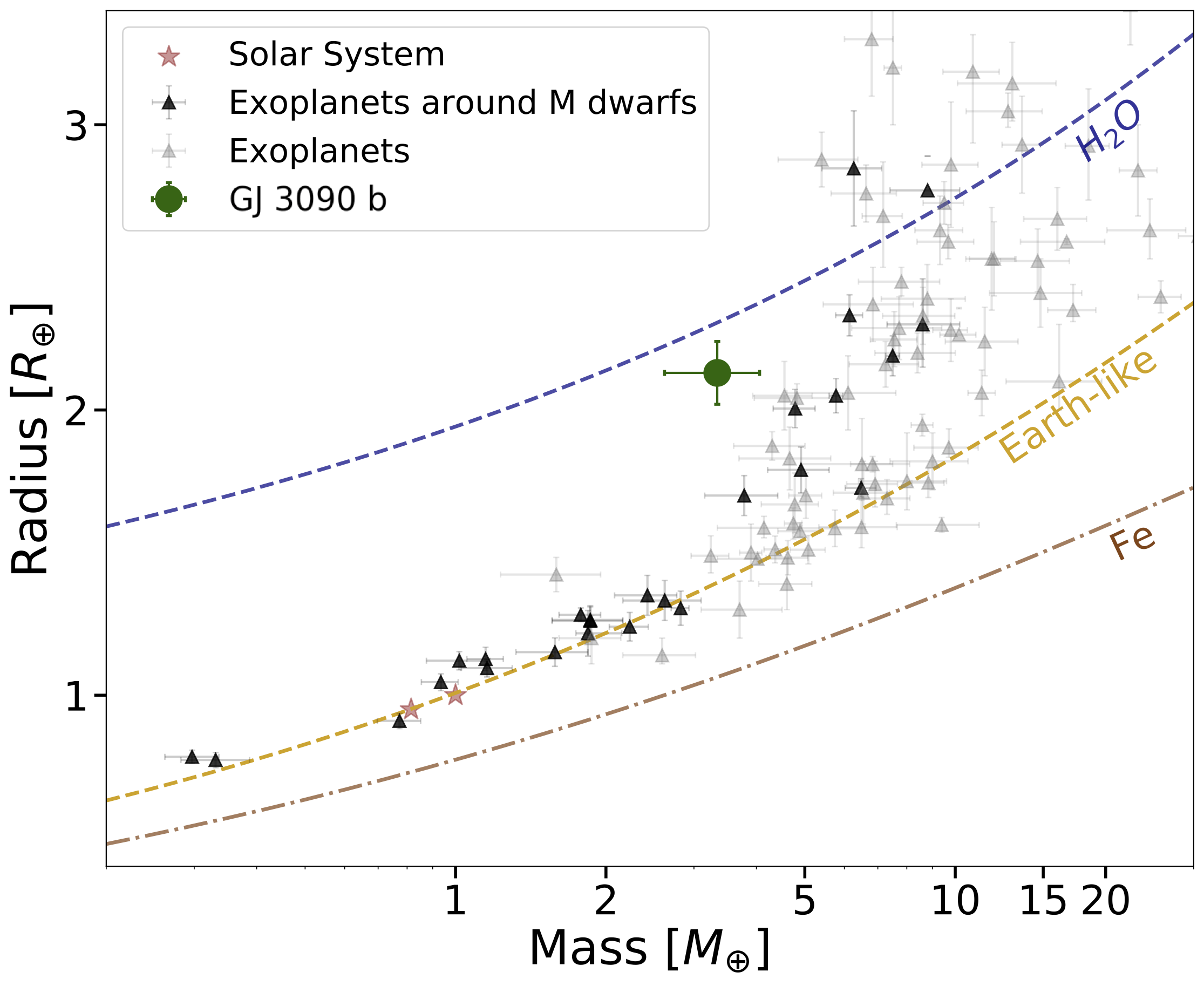}
  \end{tabular}
  \caption{Mass-radius diagram of exoplanets with an accurate mass and radius determination from the updated version of \citet[][accessible on the Data \& Analysis Center for Exoplanet DACE\protect\footnotemark]{otegi2020}. Exoplanets around M dwarfs are highlighted in black. The curves are the composition lines of iron, Earth-like, and pure water planets (at T$_{\rm eq}=693$~K).}\label{fig.mr}  
\end{figure}

With a mass of 3.34~$\pm$~0.72~\MEarth and a radius of 2.13~$\pm$~0.11~\REarth, GJ~3090~b lies at the upper edge of the transition between the populations of super-Earths and mini-Neptunes. Figure~\ref{fig.mr} shows M-R curves corresponding to planets at $T_{\rm eq}=693$~K for compositions of pure iron, an Earth-like composition (with a core mass fraction of 0.33), and compositions of pure water. For reference, we also show exoplanets with accurate and reliable mass and radius determinations \citep{otegi2020}. Figure~\ref{fig.mr} shows that planets with masses up to nearly 4~\MEarth tend to follow the Earth-like composition curve, while more massive planets have a wide diversity in radius and density. GJ~3090~b is located in the upper part of the exoplanet envelope that starts at around 4~\MEarth. Most of the exoplanets around M dwarfs are located in the upper part of the envelope, which may be due to the observational bias. These planets have a lower incoming irradiation that allows them to keep most of the H-He atmospheres. GJ~3090~b sits well above the Earth-composition curve, implying that a volatile-envelope accounts for a significant fraction of its radius. 

We modelled the interior of GJ~3090~b as the superposition of a pure-iron core, a silicate mantle, a pure-water layer, and a H-He atmosphere. The models follow the basic model structure of \cite{Dorn17}, and the equation of state (EOS) of the iron core is taken from \cite{Hakim2018}, and the EOS of the silicate-mantle is taken from \cite{Connolly09}. For water, we used the AQUA EOS from \cite{Haldemann2020}, and for the H-He envelope, the SCVH EOS \citep{Saumon1995} with an assumed protosolar composition. Precise characterisation of the internal planetary structure is very challenging because different compositions can lead to identical mass and radius \cite[][]{Rogers10,Lopez14,Dorn15,Dorn17,Lozovsky18,otegi-20-2,valencia2013}. However, Fig.~\ref{fig.mr} shows that most super-Earths follow the Earth-like composition line. In addition, chemical abundance ratios in solar metallicity stars have been found to be homogeneous within 10\% in the solar neighbourhood \cite[][]{Bedell2018}, implying that exoplanets may exhibit less compositional diversity than previously expected. It is therefore reasonable to assume an Earth-like solid interior. The low density of the planet indicates the presence of volatiles, which could be water, H-He, or a combination of these two. When an Earth-like solid interior is assumed, the water mass fraction needed to match the observed mass and radius is 55$^{+18}_{-16}$\%. On the other hand, in a water-free scenario, the required H-He mass fraction would be 4.9$^{+0.7}_{-1.3}$\%. These two values are upper limits on the water and H-He content because the planet may contain a mix of water vapour and H-He. 

\begin{figure*}[h]
\centering
  \begin{tabular}{@{}cc@{}}
    \includegraphics[width=0.95\textwidth]{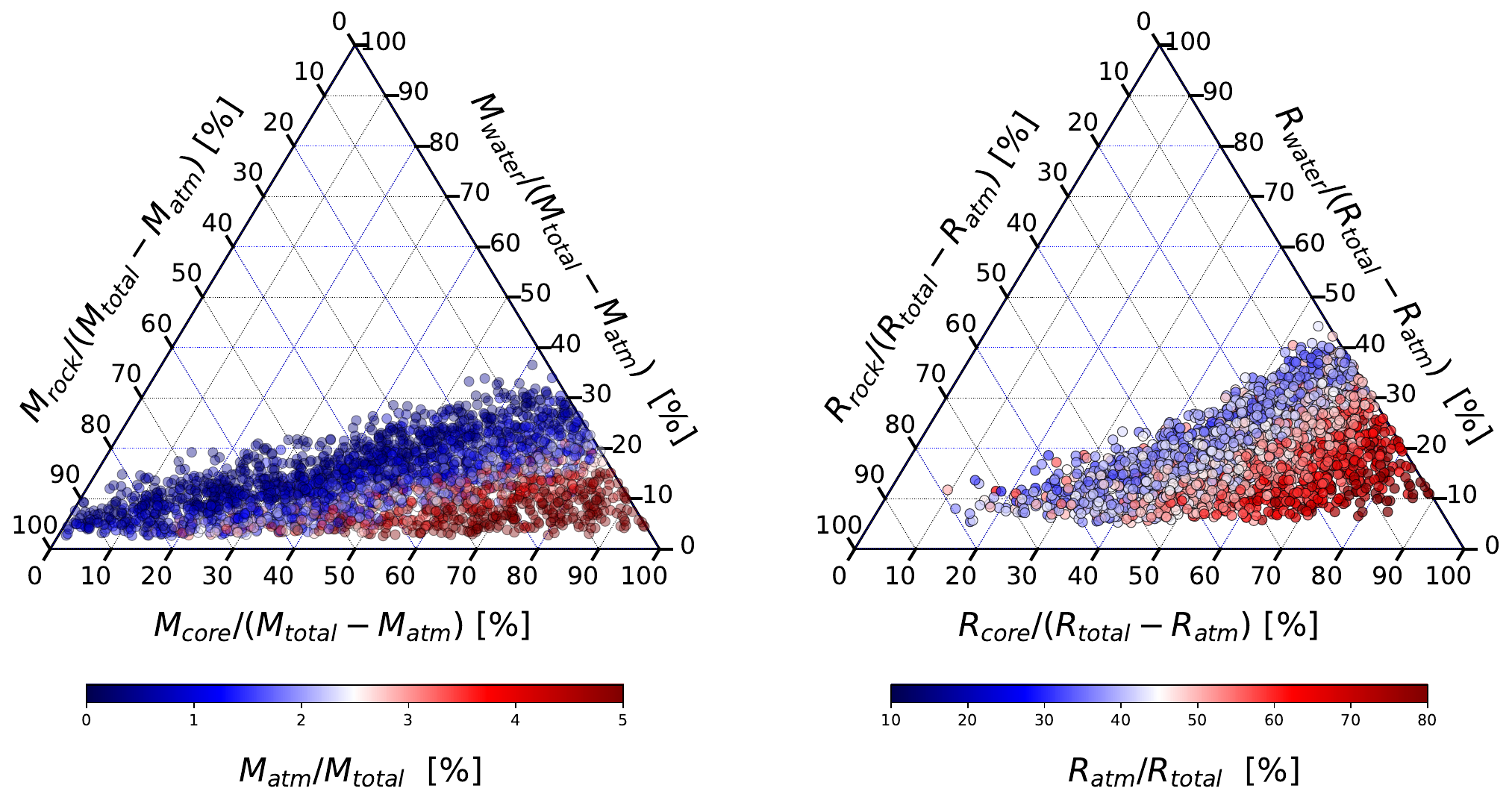}
  \end{tabular}
  \caption{Ternary diagrams of the inferred internal composition of GJ~3090~b. We show the parameter space covered by the posterior distributions in mass (left) and radius (right).}\label{fig.ternary}
\end{figure*}

Alternatively, the degeneracy can also be partly broken without assumptions on the planet abundances by using a generalised Bayesian inference method with a nested-sampling scheme. This method allows quantifying the degeneracy and correlation of the structural parameters of the planet and to estimate the most likely region in the parameter space. Figure~\ref{fig.ternary} shows ternary diagrams of the inferred composition of GJ~3090~b. The ternary diagram illustrates the composition degeneracy for exoplanets with measured mass and radius. We found a median H-He mass fraction of about 2\%, which is a lower bound because metal-enriched H-He atmospheres are more compressed and therefore increase the planetary H-He mass fraction. Formation models indeed suggest that mini-Neptunes likely form through envelope enrichment \cite[][]{Venturini2017}. We found a very strong degeneracy between the core, the silicate mantle, and the water layer, which prevents accurate estimates of the masses of these constituents. Interior models cannot distinguish between water and H-He as the source of the low-density material, and therefore we also ran three-layer models without the H$_{2}$O layer and without the H-He envelope. Table~\ref{table.interior} lists the inferred mass fractions of the core, mantle, water layer, and H-He for different models. In the water-free case, we found that the planet is 4.3\% H-He, 45\% iron, and 50.7\% rock by mass, which sets maximum limits for the atmospheric and core mass. However, a model without H-He would require 57\% water. This qualitatively agrees with previous calculations that took into account that water must be mainly in vapour and supercritical states for planets receiving an insolation above the runaway greenhouse limit, as applies for GJ~3090~b \citep{turbet2020,mousis2020,aguichine2021}. A more self-consistent estimate of the water content would be obtained by considering interactions with the rocky interior \citep{bower2019,dorn2021}. We note that the water and H-He mass fractions obtained in the three-layer models are compatible with the results obtained assuming a Earth-like solid interior. An iron-poor formation would be another possible scenario, but a volatile (either H$_{2}$/He or H$_{2}$O-dominated) layer is always required. An iron-free model gives a median atmospheric mass fraction of 0.9\%, a mantle of 84\%, and a water layer of 15\%. Because any iron added to this model would increase the H-He mass fraction and decrease the mantle mass fraction, these values set a minimum and maximum limit, respectively. We found a strong degeneracy between the core and mantle mass fractions that might be reduced with further constraints such as the stellar abundances. \\

\footnotetext{\url{https://dace.unige.ch/exoplanets/}}

\begin{table}
\renewcommand{\arraystretch}{1.4}
\setlength{\tabcolsep}{3.25pt}
  \caption{Inferred interior structure properties of GJ~3090~b.}
  \centering
\begin{tabular}{lcccc}
\hline
\hline
Constituent & 4-layer & No H-He & No H$_{2}$O & No Fe \\
\hline
$M_{\rm core}/M_{\rm total}$ & 0.44$^{+0.14}_{-0.23}$ & 0.18$^{+0.09}_{-0.07}$ & 0.45$^{+0.21}_{-0.20}$ & - \\
$M_{\rm mantle}/M_{\rm total}$ & 0.37$^{+0.21}_{-0.18}$ & 0.24$^{+0.14}_{-0.11}$ &  0.51$^{+0.23}_{-0.22}$ & 0.84$^{+0.04}_{-0.07}$  \\
$M_{\rm water}/M_{\rm total}$ & 0.17$^{+0.14}_{-0.11}$ & 0.57$^{+0.12}_{-0.13}$  & - & 0.15$^{+0.06}_{-0.04}$\\
$M_{\rm atm}/M_{\rm total}$ & 0.019$^{+0.006}_{-0.007}$ & - & 0.043$^{+0.009}_{-0.008}$ & 0.009$^{+0.005}_{-0.004}$ \smallskip\\
\hline
\end{tabular}
\label{table.interior}
\end{table}

\subsection{Coupled atmospheric and dynamical evolution}\label{sec.evaporation}

We study in this section the possible history of the atmosphere of planet b with the \texttt{JADE} code \citep{Attia21}. We refer to \citet{Attia21} for details, but briefly summarise the main ingredients here. The \texttt{JADE} code simulates the evolution of a planet over secular (gigayears) timescales under the coupled influence of complex dynamical and atmospheric mechanisms. The planet is modelled as composed of an iron core, a silicate mantle, and a H/He gaseous envelope, whose varying properties are self-consistently derived at each time step as the orbit, stellar irradiation, and inner planetary heating all evolve. The atmosphere easily erodes due to XUV-induced photoevaporation, and \texttt{JADE} calculates this mass-loss rate using analytical formulae adjusted to detailed simulations of upper atmospheric structures \citep{Salz16}. This provides a better agreement with observed mass-loss rates than the commonly used energy-limited approximation \citep[e.g.][]{Watson81,Lammer03,Erkaev07}.

\begin{figure}
  \centering
  \includegraphics[width=0.49\textwidth]{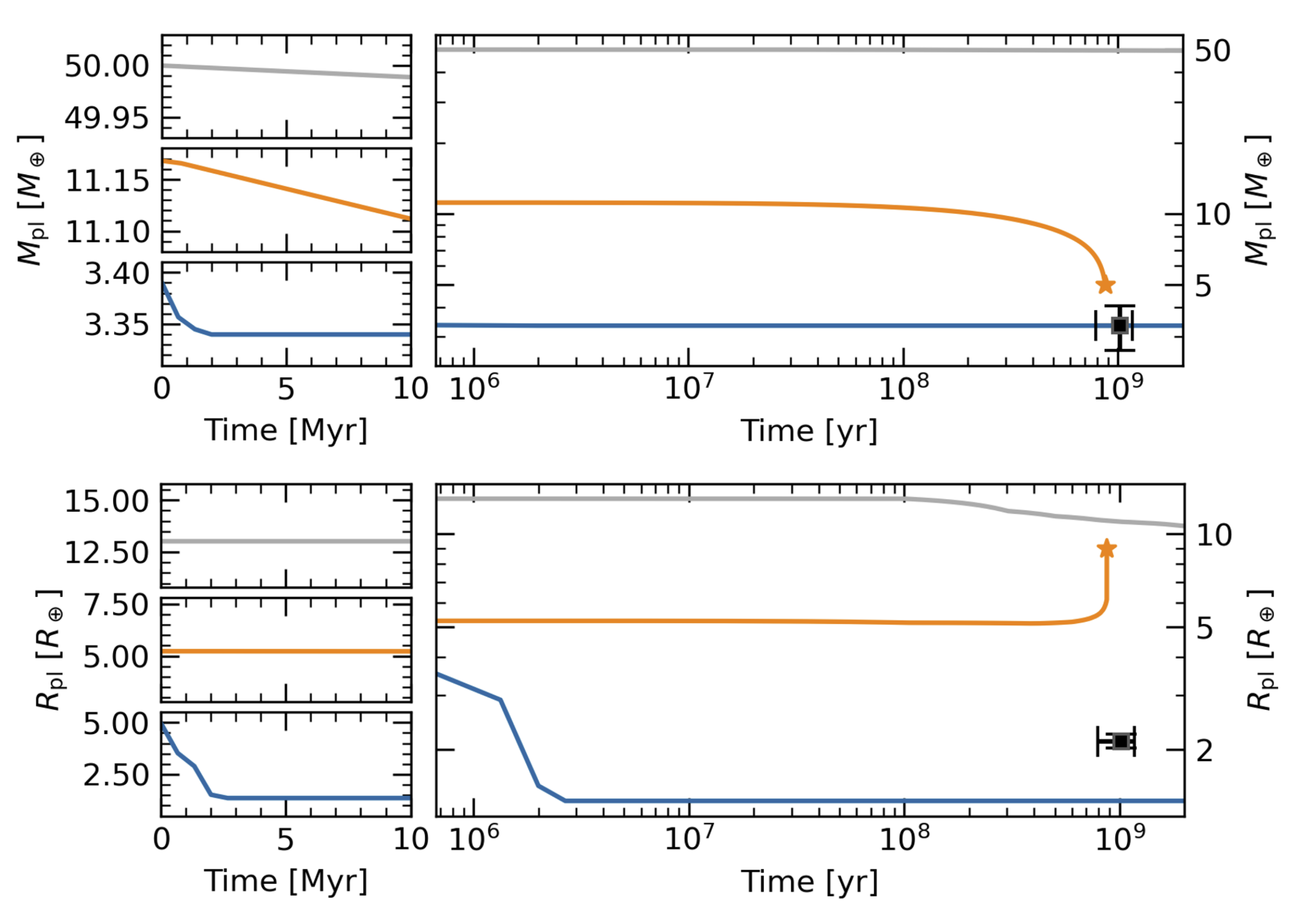}
  \caption{Evolution of three representative simulations of GJ~3090~b from the exploration of Sect.~\ref{sec.evaporation}, represented by three different colours. \textit{Top:} Evolution of the mass. \textit{Bottom:} Evolution of the radius. The inset on the left zooms into the first 10 Myr of each simulation. The star marks tidal disruption. The two black points are the observed mass and radius of GJ~3090~b at the inferred age of the system.} 
  \label{fig.evolution}
\end{figure}

We first investigated the sole impact of photoevaporation by conducting purely atmospheric simulations at a fixed orbit corresponding to the present-day values of Table \ref{table.results}. We used protosolar abundances \citep{Asplund21} for the atmospheric composition, and we accounted for the evolution of the star. The stellar irradiation, which we separated into three spectral bands (bolometric, X, and EUV), was analytically computed at each time step using the bolometric luminosity of Table \ref{table.sed} and the model of \citet{Jackson12}. We simulated the atmospheric evolution of planet b for a wide range of initial planet masses above the currently observed mass $M_{\rm pl} = 3.34$~\MEarth starting at $t=0$ from the dissipation of the protoplanetary disc. 

Within this framework, we find no initial configuration that is compatible with the current bulk properties of the planet at the inferred age of the system ($1.02^{+0.23}_{-0.15}$ Gyr, Sect.~\ref{section.stellar_parameters}). In Fig.~\ref{fig.evolution} we present the mass and radius evolution of three typical simulations that are representative of the outcomes of the exploration. The grey simulation represents a planet that is too massive to be affected by photoevaporation, showing thus nearly constant bulk properties during 2~Gyr. The orange simulation represents an intermediate-mass planet that shows a substantial inflation of the planetary radius when evaporation causes the mass to drop to $\sim 5$~\MEarth, which is typical for hot H/He-dominated low-mass atmospheres \citep[e.g.][]{Jin14,zeng2019,Gao20}. Due to this inflation, the atmosphere extends to the Roche limit of the star \citep[e.g.][]{Gu03,Jackson17} and is hence affected by tidal disruption. In this case, we ended the simulation, considering that the atmosphere is completely depleted. Finally, the blue simulation shows a low-mass planet whose atmosphere is entirely exhausted after a few million years due to photoevaporation. We highlight that only intermediate-mass planets (e.g. the orange simulation) are tidally disrupted because they have lower densities. For a given orbit, the Roche radius definition indeed provides a lower limit on the density below which stellar tidal forces exceed the planetary gravitational self-attraction.

\begin{figure}
  \hspace{-0.2cm}\includegraphics[width=0.51\textwidth]{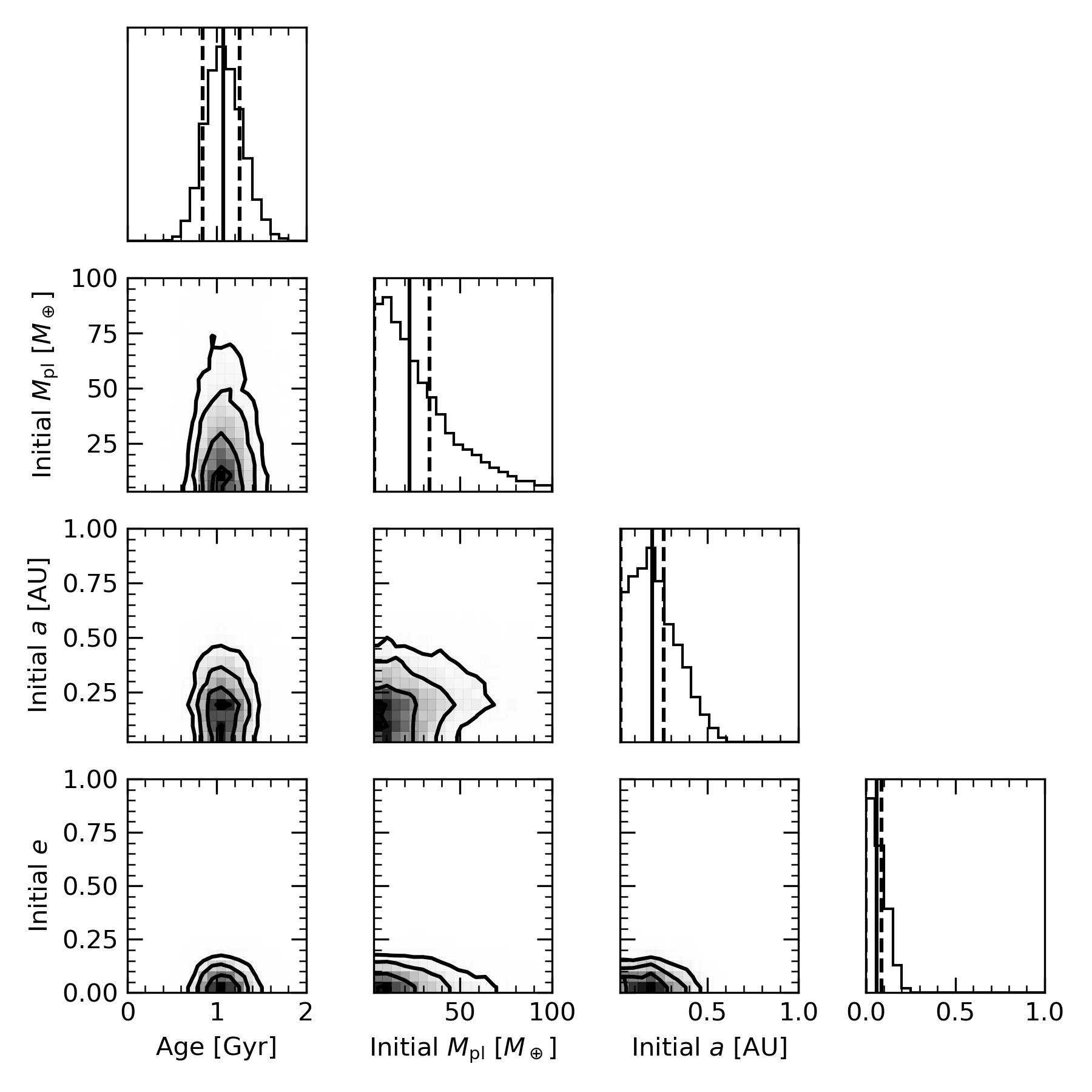}
  \caption{Corner plot of the age, initial planetary mass, initial semi-major axis, and initial eccentricity of the simulated GJ~3090~b from the exploration of Sect.~\ref{sec.evaporation}. Full vertical lines indicate the PDF median values, and dashed vertical lines indicate the $1 \sigma$ highest density intervals.} 
  \label{fig.corner}
\end{figure}

We then conducted fully coupled atmospheric and dynamical simulations by running a grid of models that covered a broad range of initial masses, semi-major axes, and eccentricities. These models also included planetary and stellar tides, planetary and stellar spin evolution, and general relativity, which are thought to be the most important dynamical forces on close-in orbits \citep[e.g.][]{Eggleton01,Mardling02}. The purpose of this exploration was to assess whether forming the planet farther away, thus escaping the intense irradiation from the young star \citep[e.g.][]{owen2017,Rogers21} before migrating only recently \citep{Bourrier18,Attia21}, circumvents the issue of radius inflation. To estimate the most likely region in the parameter space, we again employed a Bayesian inference framework (Attia et al.~\textit{in prep.)}, using \texttt{emcee} \citep{emcee} constrained by the values of Table \ref{table.results} and with non-informative priors. The jump parameters were the initial conditions and the age of the system, and the probabilities were constructed by comparing the outcomes of the simulations to the observations. The resulting corner plot \citep{ForemanMackey16} is shown in Fig.~\ref{fig.corner}. Again, no appropriate set of initial conditions is found. The most compatible configurations are low-mass planets that entirely erode after a few million years, leaving a bare core that is not consistent with the bulk properties of planet b. Planets forming with an intermediate mass are less preferred because they are still inflated and subsequently tidally disrupted (as in Fig.~\ref{fig.evolution}), even though they only recently migrated close to the star. 

This shows that whether planet b migrated earlyon or underwent a later, more complex migration for example due to interactions with planet c, it would always have lost a light H/He atmosphere in a few million years after arriving at its current location. Because the current density is also not consistent with an atmosphere-free rock-iron core, our simulations combined with the results of Sect.~\ref{sec.interior} suggest that a different atmospheric composition makes it more resilient to tidal disruption and photoevaporation. Particularly a water-enriched (i.e. high-metallicity), possibly even water-dominated atmosphere would allow for a wider range of initial masses with compact atmospheres \citep{Lopez17} because of the increased resilience of water to atmospheric escape processes. Alternatively, silicate vapour in the lower parts of the atmosphere might also result in a higher metallicity \citep{Misener22}, especially for such a close-in target. Future atmospheric characterisation of the target would help to answer this question.

\subsection{Prospects for atmospheric characterisation} 
\label{sec.atmosphere_carac}

\begin{figure*}
  \sidecaption
  \includegraphics[width=0.7\textwidth]{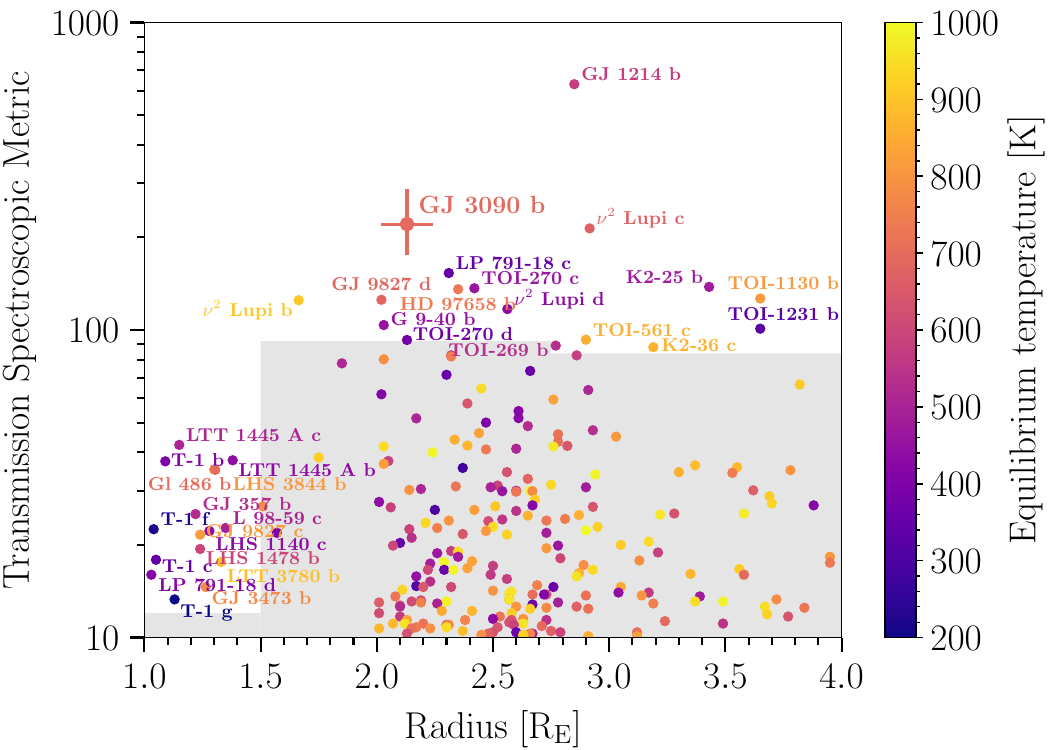}
  \caption{Transmission spectroscopy metric for planets with radii between 1 and 4~\REarth and T$_{\rm eq} < 1000$~K. Data are taken from \citet{guo2020}, updated with recent results \citep{delrez2021,burt2021,weiss2021,kemmer2020,nowak2020,trifonov2021,soto2021,cointepas2021,winters2021}. Grey areas are below the cutoffs that \citet{kempton2018} suggested for follow-up efforts. The planets above these cutoffs are labelled (T-1 is the abbreviation for Trappist-1). The symbol with error bars is GJ~3090~b. The colour indicates the equilibrium temperature of the planet, computed for zero albedo and full day-night heat redistribution.} \label{fig.TSM}
\end{figure*}

Because it is very close to us (22~pc) and orbits a relatively low-mass M2V star, GJ~3090~b is one of the best mini-Neptune-sized planets for a characterisation of its atmosphere to date. The transmission spectroscopy metric \citep[TSM,][]{kempton2018} of GJ~3090~b is $221^{+66}_{-46}$, which means that only one other of the known mini-Neptunes has an atmosphere that can be better characterised through transit spectroscopy (Fig.~\ref{fig.TSM}). Below, we further evaluate and discuss the prospects for characterising the true nature of the atmosphere of GJ~3090~b.

As shown in Sect.~\ref{sec.interior}, theoretical mass-radius relations for planets predict that GJ~3090~b has a volatile-rich envelope that is dominated in mass by either hydrogen and helium, or water. Moreover and as shown in Sect.~\ref{sec.evaporation}, stability to atmospheric escape requirements suggest that the atmosphere of GJ~3090~b is likely to have super-solar metallicity. It should therefore have a high water-mass fraction. A high metallicity is consistent with the empirical mass-metallicity trend of Solar System planets (see \citealt{Wakeford2020} and references therein). In-depth analyses of HST/WFC3 transit spectra of some of the least massive mini-Neptunes (e.g. GJ~1214~b and K2-18~b; see \citealt{Kreidberg2014,Benneke2019b,Tsiaras2019}) provide tentative evidence that their metallicity is super-solar \citep{Charnay2015b,Gao2018,Charnay2021}, which may -- at least for the population of mini-Neptunes most comparable in mass to GJ~3090~b -- support this empirical trend.

\begin{figure}
  \hspace{-0.4cm}\includegraphics[width=0.54\textwidth]{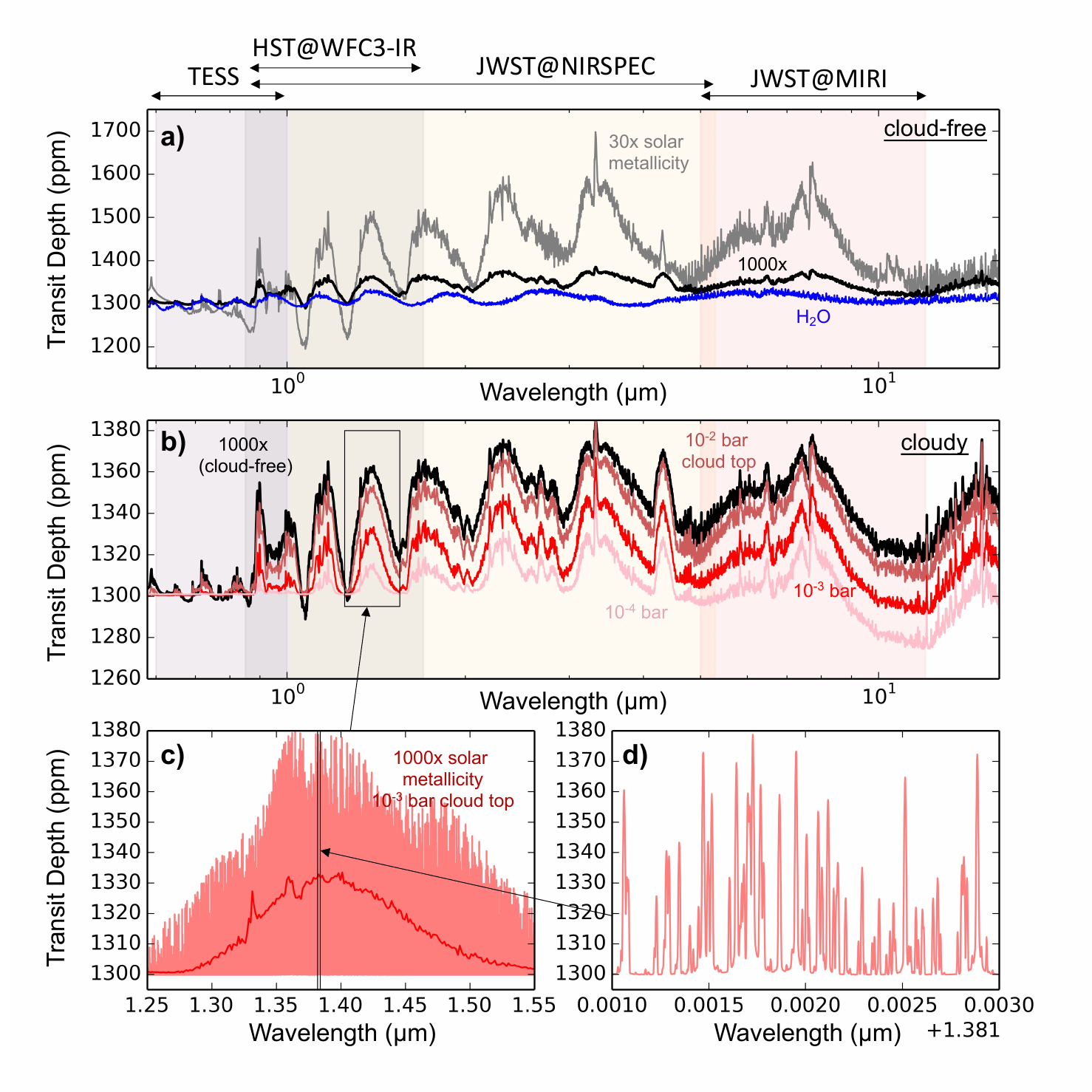}
  \caption{Synthetic transit spectra of GJ~3090~b for different (hypothetical) atmospheric compositions. (a) Transit spectra for cloud-free atmospheres (30x solar metallicity, 1000x solar metallicity, and pure water vapour). (b) Transit spectra for cloudy atmospheres with 1000x solar metallicity with cloud layers at different altitudes (10$^{-2}$, 10$^{-3}$, 10$^{-4}$~bar, and cloud-free). (c-d) Two levels of zoom into the water-absorption lines (in the 1.4~$\mu$m water-absorption band) at high spectral resolution for a 1000x solar metallicity atmosphere with a cloud top at 10$^{-3}$~bar. The spectral ranges of TESS, HST/WFC3, JWST/NIRSPEC, and JWST/MIRI have been added for reference.} \label{fig.transit_spectra}
\end{figure}

To further evaluate the feasibility of characterising the atmosphere of GJ~3090~b, we computed synthetic transit spectra (summarised in Fig.~\ref{fig.transit_spectra}) from visible to infrared wavelengths (0.5-16~$\mu$m) at low and high spectral resolutions for different atmospheric scenarios (low metallicity, high metallicity, pure water vapour, cloud-free, and cloudy). To construct these spectra, we proceeded as follows. (1) Following \citet{Villanueva2018}, we calculated the pressure-temperature profiles using the prescriptions of \citet{Parmentier2014} and assuming chemical equilibrium for the atmospheric composition \citep{Mbarek2016,Kempton2017}. (2) Transmission spectra were computed with petitRADTRANS \citep{Molliere2019} using the temperature, pressure, and mixing ratio vertical profiles. For simplicity, we used 1D atmospheric models. Future works should revisit these estimates with 3D atmospheric models, first to more realistically simulate the position of clouds (see e.g. \citealt{Charnay2015b,Charnay2021}), and then to account for the 3D transit geometry, which is particularly relevant for low-mass highly irradiated planets with a potentially low mean molecular weight atmosphere \citep{Caldas2019,Pluriel2020,Macdonald2022,Wardenier2022}.

As expected from previous studies \citep{Morley2013,Morley:2015,Charnay2015b,Greene:2016,Molliere2017,Kawashima2018}, the amplitude of the transmission spectra is highly dependent on metallicity (for the explored range of metallicity, the higher the metallicity, the higher the average molecular weight of the atmosphere, the lower the scale height, and thus the lower the amplitude of the transit spectroscopy signal) and on the presence and altitude of the cloud (top) layer. The simulated transit depth amplitudes range from $\sim$~200~ppm (30$\times$ solar metallicity, cloud-free) down to 30~ppm (1000$\times$ solar metallicity, cloudy atmosphere). Based on previous observations \citep{Kreidberg2014} and models \citep{Charnay2015b,Gao2020} of irradiated mini-Neptunes, clouds are very likely to be present in the atmosphere of GJ~3090~b, and thus are likely to mute the transmission spectrum.

Scaling from GJ~1214~b observations \citep{Kreidberg2014}, we estimate that a $\sim$~25~hours HST program using the G141 grism of the WFC3-IR instrument would provide a $\sim$~50~ppm transit depth uncertainty (around 1.4~$\mu$m, for R=70), which is similar to that of \citet{Kreidberg2014}. For a given atmospheric composition, however, the expected amplitude of the near-infrared transit spectroscopy signal is lower for GJ~3090~b than for GJ~1214~b by a factor of $\sim$~7. This factor is mostly justified by its radius, which is 1.33 times smaller. Such an HST/WFC3-IR observation program would be able to characterise an atmosphere of low to moderate metallicity, and only in the absence of high-altitude clouds and/or hazes.

\begin{figure*}
  \centering
  \includegraphics[width=0.85\textwidth]{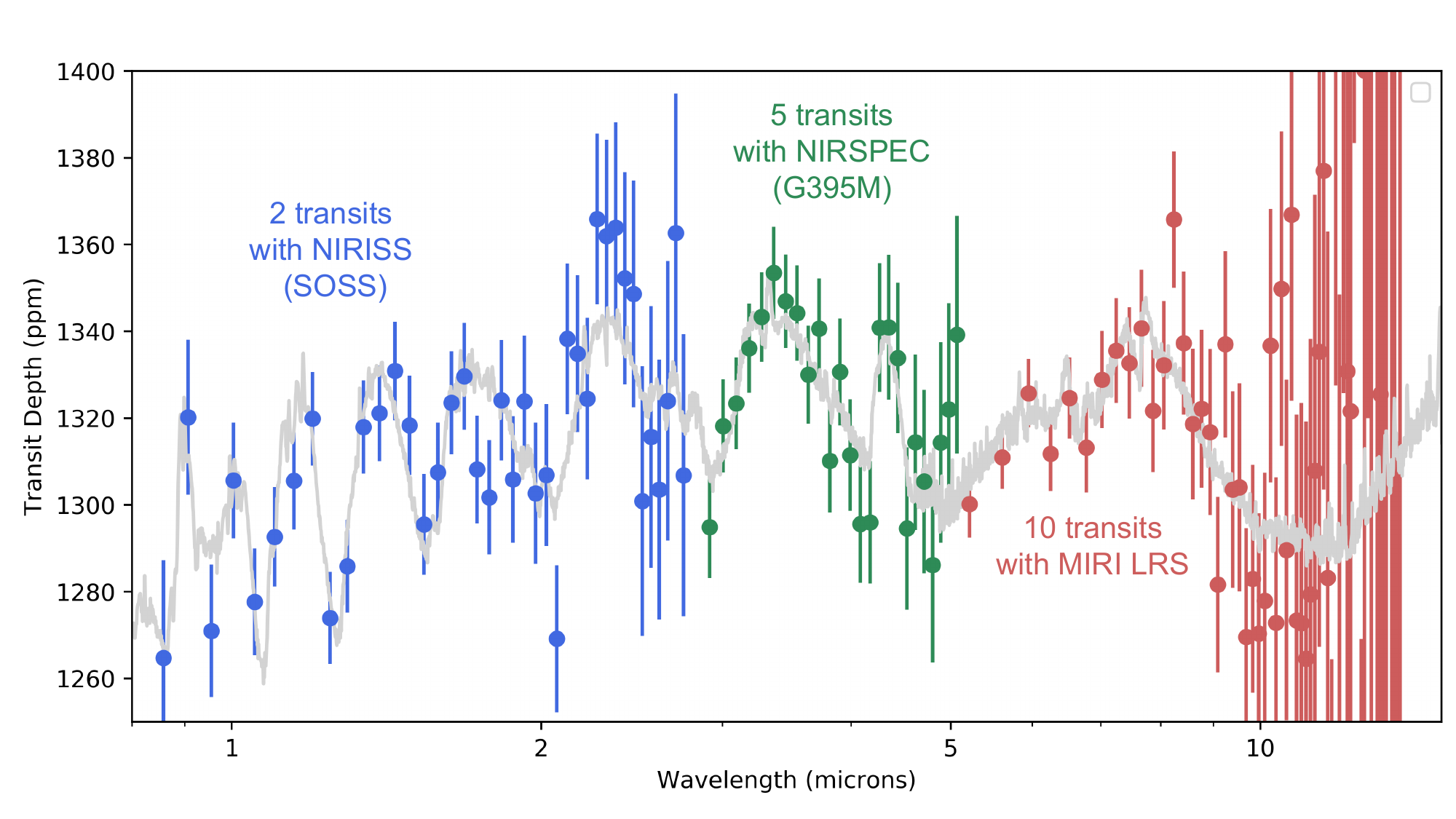}
  \caption{Synthetic transit spectra of GJ~3090~b. The estimated uncertainties are calculated with the JWST ETC PandExo \citep{Batalha2017} using NIRISS (SOSS; blue) with the NIRSPec (395M grism; green) and with MIRI (LRS; red) instrument modes of the JWST. This calculation assumes a 1000x solar metallicity atmosphere (with a cloud top at 1~millibar). The initial modelled transit spectrum is plotted in grey.} \label{fig.transit_spectra_jwst_pandexo}
\end{figure*}

To make a step further in characterising the atmosphere of GJ~3090~b, JWST \citep{gardner2006} observations are a very promising path. Using the JWST ETC PandExo \citep{Batalha2017}, we evaluated the detectability of the atmosphere of GJ~3090~b with NIRISS, NIRSPec, and MIRI-LRS for the range of previously discussed atmospheric scenarios (shown in Fig.~\ref{fig.transit_spectra}). Figure~\ref{fig.transit_spectra_jwst_pandexo} shows an illustrative result for a 1000x solar metallicity atmosphere with a cloud top at 1~millibar. We find that NIRISS observations are the most promising for this range of simulated atmospheres. GJ~3090 is too bright to use NIRSPec-prism, and its rapid brightness decrease in the mid-infrared (compared to a later-type star such as GJ~1214) makes MIRI observations expensive. A reconnaissance observation program of two transits with NIRISS single-object slitless spectroscopy (SOSS) -- fewer than 10 hours of JWST time, after taking overheads into account -- could indeed characterise the atmosphere of GJ~3090~b even if its metallicity is high and it has high clouds. The spectral coverage of transit spectra could be extended to infrared wavelengths with NIRSPec-G395H/M observations, but this would require two to three times more transits to give a similar signal-to-noise ratio. A combination of transit spectra obtained with NIRISS (SOSS) and NIRSPec (G395H/M) has indeed been shown to be the optimal strategy for bright targets such as GJ~3090 \citep{Batalha2017b,Batalha2018}. None of the high-metallicity scenarios can be characterised with MIRI-LRS in transit spectroscopy with a reasonable number of observation hours. However, and given the expected equilibrium temperature of GJ~3090~b, thermal emission might be searched for with MIRI (in LRS mode, or alternatively, using a combination of filters) in a handful of secondary eclipses.

\begin{figure}
  \centering
  \includegraphics[width=0.49\textwidth]{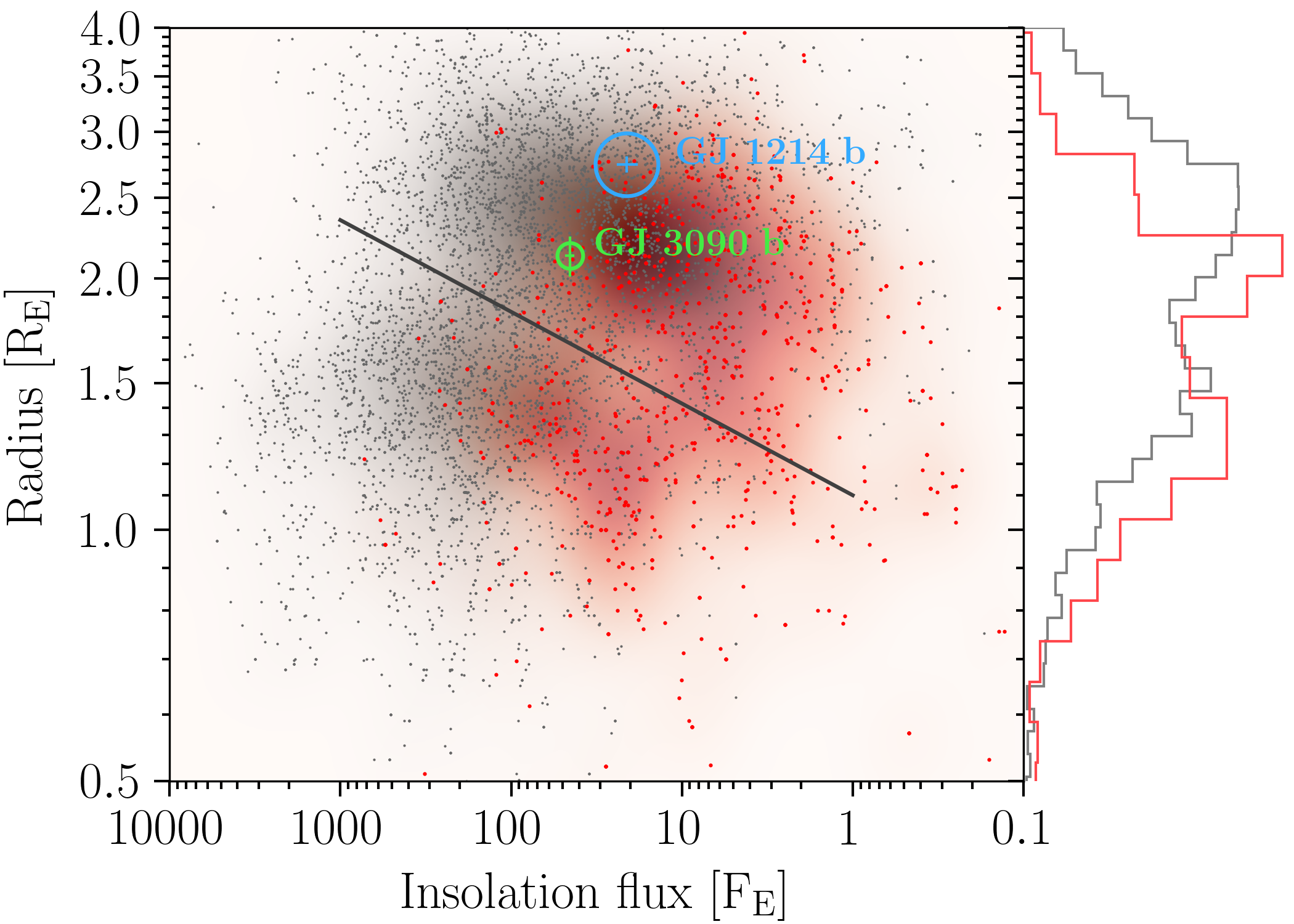}
    \caption{Planet radius vs insolation flux in Earth units and logarithmic scale. Error bars mark the position of GJ~3090~b (green) and GJ~1214~b (blue). The size of the symbols is proportional to the planet mass. Grey dots are planets listed in the NASA Exoplanet Archive (\url{https://exoplanetarchive.ipac.caltech.edu/}) with radius uncertainties smaller than 20\%. A Gaussian kernel density estimate is shown in different background shades of grey. A subset of planets whose stars have a $T_{\mathrm{eff}} < 4000~K$ is shown as red dots, and its Gaussian kernel density estimate is shown in different background shades of red. The histogram of the entire sample (grey) and the subset (red) is shown right of the panel. The black line shows the scaling relation for the radius valley predicted by the photoevaporation model of \citet{lopez2018}. The two main overdensities are the mini-Neptunes (above the line) and the super-Earths (below the line). The scarcity of planets in the upper left part of the diagram is the so-called photoevaporation desert, hot-Neptune desert, or hot-super-Earth desert \citep{lundkvist2016}.}\label{fig.RadiusIrradiation}
\end{figure}

\medskip
GJ~3090~b is highly complementary to the benchmark GJ~1214~b \citep{charbonneau2009,cloutier2021}. It is almost half as massive and falls in a different region of the planetary radius-insolation diagram (Fig.~\ref{fig.RadiusIrradiation}), immediately above the radius valley \citep{fulton2017,cloutier2020}. Therefore, GJ~3090~b it is an excellent probe of the edge of the transition between super-Earths and mini-Neptunes.  

\begin{acknowledgements}
Funding for the TESS mission is provided by NASA's Science Mission Directorate. We acknowledge the use of public TESS data from pipelines at the TESS Science Office and at the TESS Science Processing Operations Center. This research has made use of the Exoplanet Follow-up Observation Program website, which is operated by the California Institute of Technology, under contract with the National Aeronautics and Space Administration under the Exoplanet Exploration Program. Resources supporting this work were provided by the NASA High-End Computing (HEC) Program through the NASA Advanced Supercomputing (NAS) Division at Ames Research Center for the production of the SPOC data products. This paper includes data collected by the TESS mission that are publicly available from the Mikulski Archive for Space Telescopes (MAST). 

We are grateful to the ESO/La Silla staff for their support. 

This work has made use of data from the European Space Agency (ESA) mission {\it Gaia} (\url{https://www.cosmos.esa.int/gaia}), processed by the {\it Gaia} Data Processing and Analysis Consortium (DPAC, \url{https://www.cosmos.esa.int/web/gaia/dpac/consortium}). Funding for the DPAC has been provided by national institutions, in particular the institutions participating in the {\it Gaia} Multilateral Agreement.

This work makes use of observations from the LCOGT network. Part of the LCOGT telescope time was granted by NOIRLab through the Mid-Scale Innovations Program (MSIP). MSIP is funded by NSF.
   
Based in part on observations obtained at the Southern Astrophysical Research (SOAR) telescope, which is a joint project of the Minist\'{e}rio da Ci\^{e}ncia, Tecnologia e Inova\c{c}\~{o}es (MCTI/LNA) do Brasil, the US National Science Foundation’s NOIRLab, the University of North Carolina at Chapel Hill (UNC), and Michigan State University (MSU).

This work has been supported by a grant from Labex OSUG@2020 (Investissements d'avenir -- ANR10 LABX56).

Based on data collected under the ExTrA project at the ESO La Silla Paranal Observatory. ExTrA is a project of Institut de Plan\'etologie et d'Astrophysique de Grenoble (IPAG/CNRS/UGA), funded by the European Research Council under the ERC Grant Agreement n. 337591-ExTrA.

MA, VB and MT's work has been carried out in the frame of the National Centre for Competence in Research PlanetS supported by the Swiss National Science Foundation (SNSF). MA and VB acknowledge funding from the European Research Council (ERC) under the European Union’s Horizon 2020 research and innovation programme (project {\sc Spice Dune}; grant agreement No 947634). M.T. thanks the Gruber Foundation for its support to this research.

N. A.-D. acknowledges the support of FONDECYT project 3180063.
   
\end{acknowledgements}

\bibliographystyle{aa}
\bibliography{TOI-177}

\begin{appendix}
\section{Additional figures and tables}

\begin{figure*}
  \centering
  \includegraphics[width=0.33\textwidth]{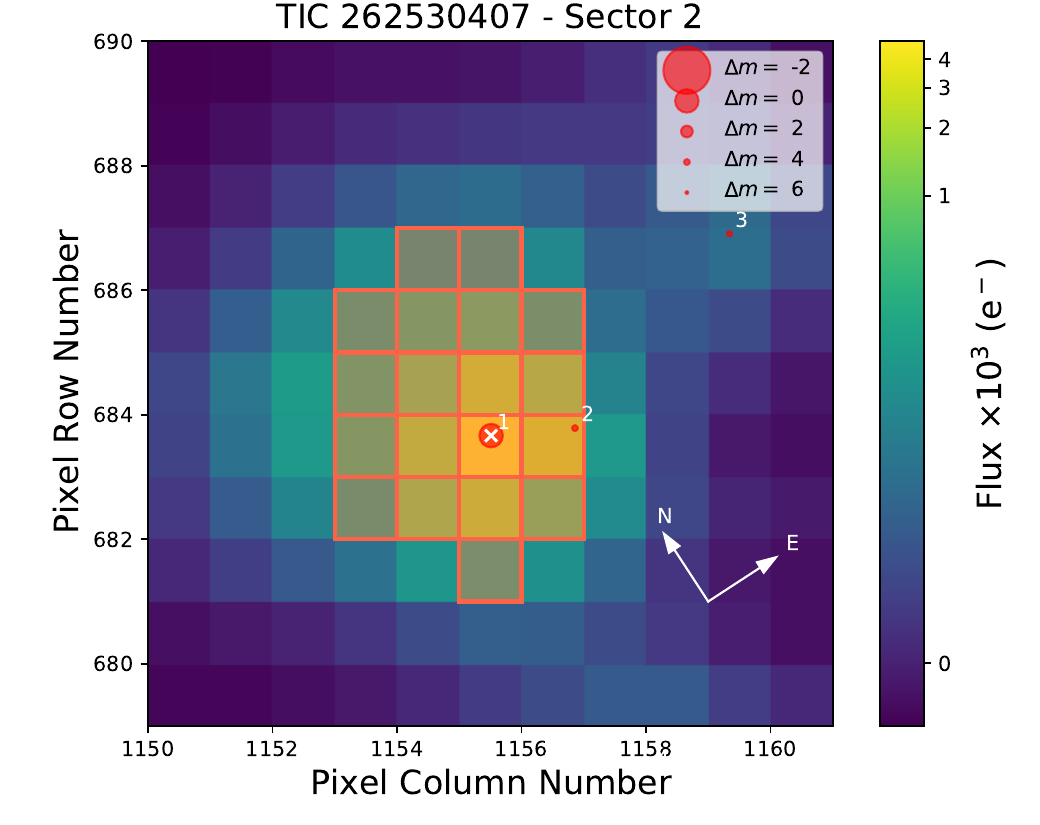}
  \includegraphics[width=0.33\textwidth]{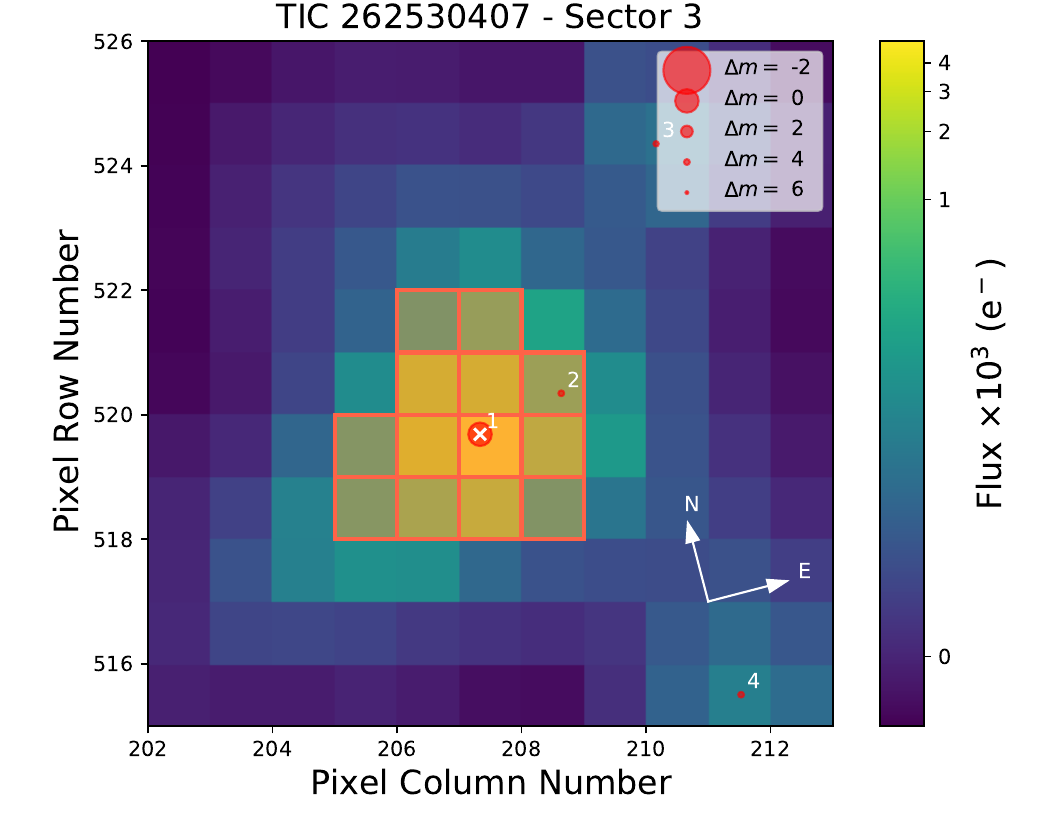}
  \includegraphics[width=0.33\textwidth]{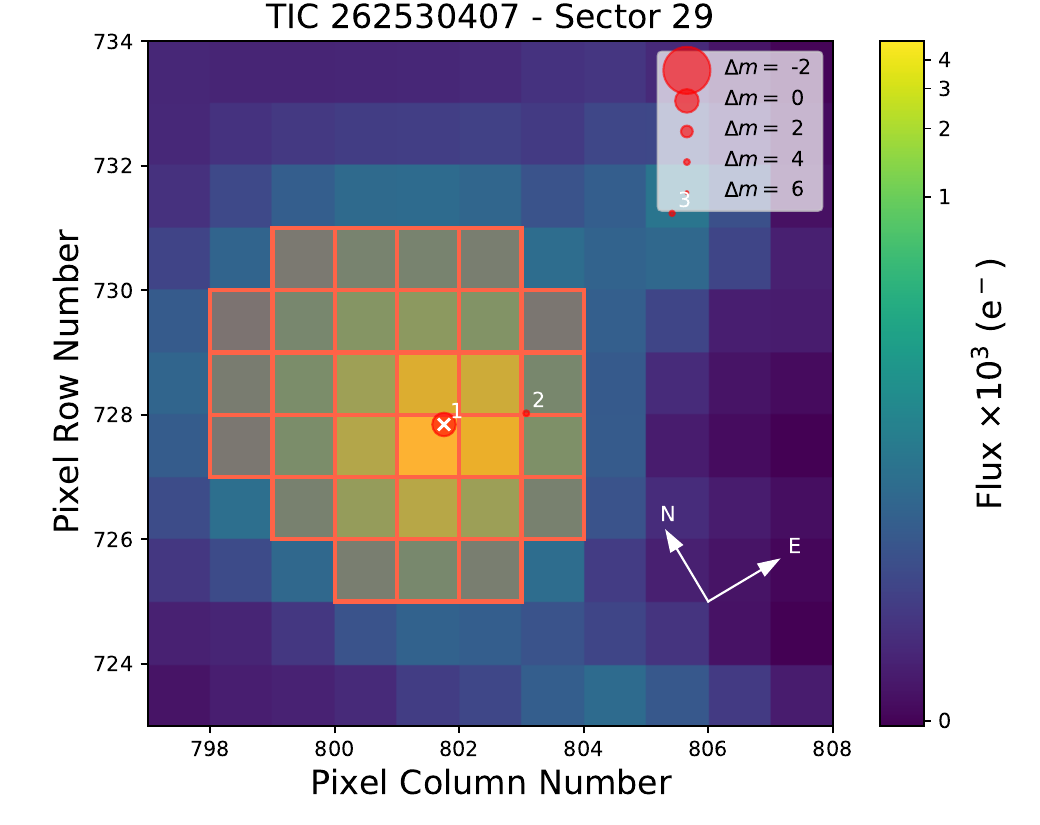}
  \caption{TESS target pixel file image of GJ~3090 in Sectors 2, 3, and 29 \citep[created with {\sc \tt tpfplotter},][]{aller2020}. The electron counts are colour-coded. The pixels highlighted in red are used for the simple aperture photometry. The positions of the stars in the Gaia DR2 \citep{gaia2018} are indicated with red circles (and labelled with numbers according to the distance to the main target in the aperture labelled "1"), their size is proportional to the Gaia DR2 magnitudes.} \label{fig.tpfplotter}
\end{figure*}

\begin{table*}
\setlength{\tabcolsep}{4pt}
\scriptsize
  \caption{Parameters measured on HARPS spectra of GJ~3090.}\label{table.rv}
\begin{tabular}{lccccccccccccccc}
\hline
\hline
 & &  & CCF & CCF & CCF &  \\
Time & RV &  $\pm1~\sigma$ & FWHM & contrast & bisector span & H$\alpha$ & $\sigma_{\rm H\alpha}$ & H$\beta$ & $\sigma_{\rm H\beta}$ & H$\gamma$ & $\sigma_{\rm H\gamma}$ & NaD & $\sigma_{\rm NaD}$ & S$_{\rm HK}$ & $\sigma_{\rm S_{\rm HK}}$\\
$[{\rm BJD}]$ & [m\,s$^{-1}$] & [m\,s$^{-1}$] & [km\,s$^{-1}$] & [\%] & [m\,s$^{-1}$]\\
\hline
2458451.685318 & 17400.46 & 1.51 & 3.35795 & 15.88118 & -0.00033 & 0.07895 & 0.00016 & 0.06984 & 0.00038 & 0.1392 & 0.0011 & 0.01584 & 0.00012 & 2.910 & 0.070 \\
2458452.659111 & 17399.92 & 1.42 & 3.35402 & 15.89489 & 0.00470 & 0.07954 & 0.00015 & 0.06970 & 0.00036 & 0.1392 & 0.0010 & 0.01613 & 0.00011 & 3.035 & 0.068 \\
2458453.659986 & 17400.81 & 1.57 & 3.35787 & 15.93346 & 0.00657 & 0.08006 & 0.00017 & 0.07260 & 0.00040 & 0.1466 & 0.0012 & 0.01603 & 0.00013 & 2.981 & 0.073 \\
2458454.611072 & 17412.42 & 2.18 & 3.37451 & 16.24279 & -0.00216 & 0.08077 & 0.00025 & 0.07143 & 0.00050 & 0.1449 & 0.0013 & 0.01645 & 0.00020 & 2.995 & 0.074 \\
2458455.689244 & 17415.93 & 2.16 & 3.36805 & 16.06612 & -0.00762 & 0.07892 & 0.00024 & 0.06795 & 0.00051 & 0.1419 & 0.0014 & 0.01592 & 0.00019 & 3.012 & 0.087 \\
2458501.556874 & 17409.55 & 2.29 & 3.37032 & 16.05940 & 0.00948 & 0.08461 & 0.00026 & 0.07695 & 0.00059 & 0.1520 & 0.0016 & 0.01693 & 0.00021 & 3.163 & 0.097 \\
2458501.568610 & 17412.18 & 2.16 & 3.36481 & 16.03897 & -0.00344 & 0.08469 & 0.00024 & 0.07623 & 0.00056 & 0.1502 & 0.0016 & 0.01654 & 0.00019 & 3.069 & 0.096 \\
2458502.554355 & 17404.14 & 1.67 & 3.35973 & 15.98065 & 0.00704 & 0.08332 & 0.00019 & 0.07644 & 0.00044 & 0.1557 & 0.0013 & 0.01670 & 0.00014 & 3.038 & 0.081 \\
2458502.566287 & 17403.38 & 1.68 & 3.35793 & 15.93048 & -0.00297 & 0.08305 & 0.00019 & 0.07536 & 0.00045 & 0.1494 & 0.0013 & 0.01641 & 0.00014 & 3.019 & 0.086 \\
2458503.528851 & 17403.06 & 1.64 & 3.35135 & 15.90108 & -0.00137 & 0.08414 & 0.00018 & 0.07779 & 0.00044 & 0.1598 & 0.0013 & 0.01711 & 0.00014 & 2.941 & 0.079 \\
2458504.530673 & 17400.19 & 1.70 & 3.36297 & 16.01077 & 0.00884 & 0.07866 & 0.00019 & 0.06956 & 0.00041 & 0.1414 & 0.0012 & 0.01605 & 0.00014 & 2.700 & 0.070 \\
2458505.528804 & 17401.71 & 1.67 & 3.36931 & 16.07097 & -0.00459 & 0.07771 & 0.00019 & 0.06757 & 0.00039 & 0.1402 & 0.0011 & 0.01591 & 0.00014 & 2.748 & 0.066 \\
2458505.540007 & 17401.50 & 1.55 & 3.36802 & 16.11554 & 0.00304 & 0.07785 & 0.00017 & 0.06848 & 0.00037 & 0.1420 & 0.0010 & 0.01582 & 0.00013 & 2.762 & 0.062 \\
2458506.533579 & 17409.04 & 1.49 & 3.36660 & 16.08706 & 0.00080 & 0.07702 & 0.00017 & 0.06586 & 0.00035 & 0.1371 & 0.0010 & 0.01553 & 0.00012 & 2.772 & 0.062 \\
2458506.545199 & 17410.67 & 1.46 & 3.36419 & 16.07681 & -0.00320 & 0.07717 & 0.00016 & 0.06739 & 0.00035 & 0.1368 & 0.0010 & 0.01556 & 0.00012 & 2.776 & 0.064 \\
2458507.541294 & 17415.10 & 1.10 & 3.36956 & 16.13506 & 0.00257 & 0.07554 & 0.00012 & 0.06448 & 0.00025 & 0.1311 & 0.0007 & 0.01536 & 0.00009 & 2.688 & 0.037 \\
2458515.529434 & 17407.78 & 2.52 & 3.36153 & 16.04329 & 0.00875 & 0.08690 & 0.00028 & 0.08175 & 0.00067 & 0.1654 & 0.0021 & 0.01738 & 0.00024 & 3.373 & 0.160 \\
2458516.534667 & 17401.38 & 1.73 & 3.36335 & 16.07780 & -0.00141 & 0.08495 & 0.00019 & 0.07876 & 0.00046 & 0.1573 & 0.0014 & 0.01664 & 0.00015 & 3.286 & 0.104 \\
2458517.533062 & 17408.08 & 1.93 & 3.36611 & 16.22144 & -0.00005 & 0.08836 & 0.00022 & 0.08685 & 0.00051 & 0.1720 & 0.0015 & 0.01780 & 0.00017 & 3.476 & 0.106 \\
2458518.538737 & 17411.35 & 3.34 & 3.35353 & 16.03431 & -0.01362 & 0.08586 & 0.00036 & 0.07910 & 0.00090 & 0.1682 & 0.0033 & 0.01774 & 0.00034 & 3.187 & 0.323 \\
2458520.545345 & 17418.12 & 2.36 & 3.36035 & 15.98638 & 0.00927 & 0.08614 & 0.00026 & 0.08452 & 0.00065 & 0.1803 & 0.0020 & 0.01776 & 0.00022 & 3.326 & 0.149 \\
2458521.528513 & 17410.08 & 2.48 & 3.35432 & 15.88085 & 0.00351 & 0.09670 & 0.00029 & 0.10741 & 0.00075 & 0.2247 & 0.0023 & 0.02109 & 0.00024 & 4.316 & 0.153 \\
2458522.526715 & 17402.21 & 2.04 & 3.36694 & 15.89182 & 0.00245 & 0.07871 & 0.00022 & 0.07066 & 0.00052 & 0.1476 & 0.0017 & 0.01618 & 0.00018 & 2.833 & 0.141 \\
2458528.520417 & 17414.55 & 2.89 & 3.38350 & 15.99426 & -0.00201 & 0.08100 & 0.00031 & 0.07312 & 0.00071 & 0.1458 & 0.0022 & 0.01694 & 0.00028 & 2.916 & 0.185 \\
2458530.523479 & 17411.69 & 2.09 & 3.36487 & 16.02049 & 0.00570 & 0.08518 & 0.00023 & 0.08002 & 0.00056 & 0.1662 & 0.0018 & 0.01740 & 0.00019 & 3.212 & 0.142 \\
2458531.518947 & 17408.19 & 1.73 & 3.36275 & 16.02831 & -0.00712 & 0.08503 & 0.00019 & 0.07717 & 0.00046 & 0.1559 & 0.0015 & 0.01726 & 0.00015 & 3.081 & 0.116 \\
2458537.519705 & 17410.79 & 2.21 & 3.35849 & 16.02056 & -0.00164 & 0.08591 & 0.00024 & 0.07931 & 0.00060 & 0.1595 & 0.0021 & 0.01691 & 0.00020 & 3.381 & 0.178 \\
2458538.517565 & 17407.23 & 2.41 & 3.36078 & 16.03019 & -0.00210 & 0.08303 & 0.00026 & 0.07459 & 0.00063 & 0.1424 & 0.0022 & 0.01642 & 0.00022 & 3.021 & 0.189 \\
2458539.504916 & 17395.49 & 1.75 & 3.36987 & 15.91665 & 0.00852 & 0.08257 & 0.00020 & 0.07377 & 0.00045 & 0.1451 & 0.0015 & 0.01648 & 0.00015 & 2.767 & 0.103 \\
2458541.502377 & 17395.66 & 2.52 & 3.37507 & 15.88192 & -0.00129 & 0.07896 & 0.00028 & 0.06899 & 0.00058 & 0.1346 & 0.0018 & 0.01641 & 0.00023 & 2.467 & 0.126 \\
2458741.680613 & 17412.50 & 1.52 & 3.39381 & 16.14142 & 0.00439 & 0.08017 & 0.00018 & 0.07138 & 0.00036 & 0.1441 & 0.0010 & 0.01652 & 0.00012 & 3.198 & 0.038 \\
2458744.791074 & 17404.74 & 1.92 & 3.38396 & 16.23178 & -0.01427 & 0.08456 & 0.00023 & 0.07694 & 0.00047 & 0.1556 & 0.0014 & 0.01757 & 0.00015 & 3.634 & 0.065 \\
2458746.611509 & 17421.12 & 3.99 & 3.38644 & 16.22588 & 0.01386 & 0.08533 & 0.00045 & 0.07616 & 0.00098 & 0.1570 & 0.0031 & 0.01671 & 0.00039 & 3.284 & 0.197 \\
2458748.654565 & 17427.93 & 2.07 & 3.38101 & 16.26426 & 0.00572 & 0.08422 & 0.00024 & 0.07616 & 0.00052 & 0.1461 & 0.0015 & 0.01644 & 0.00017 & 3.229 & 0.076 \\
2458754.630341 & 17390.93 & 2.21 & 3.38283 & 16.14074 & 0.00072 & 0.08424 & 0.00026 & 0.07554 & 0.00055 & 0.1524 & 0.0016 & 0.01680 & 0.00019 & 3.098 & 0.081 \\
2458756.690323 & 17404.88 & 2.39 & 3.39395 & 16.08237 & 0.01304 & 0.08697 & 0.00029 & 0.08087 & 0.00060 & 0.1660 & 0.0018 & 0.01728 & 0.00021 & 3.247 & 0.087 \\
2458757.604314 & 17407.92 & 2.49 & 3.38763 & 16.11348 & 0.00621 & 0.08342 & 0.00029 & 0.07718 & 0.00061 & 0.1597 & 0.0019 & 0.01697 & 0.00022 & 3.295 & 0.100 \\
2458760.714200 & 17416.04 & 2.28 & 3.38513 & 16.06580 & 0.00450 & 0.08022 & 0.00026 & 0.07023 & 0.00055 & 0.1425 & 0.0016 & 0.01612 & 0.00019 & 2.972 & 0.090 \\
2458760.725612 & 17410.24 & 2.10 & 3.38636 & 16.11989 & 0.00107 & 0.08019 & 0.00024 & 0.06915 & 0.00050 & 0.1381 & 0.0015 & 0.01615 & 0.00018 & 2.912 & 0.082 \\
2458761.865381 & 17408.40 & 1.91 & 3.36370 & 15.93534 & -0.00729 & 0.08335 & 0.00022 & 0.07495 & 0.00052 & 0.1470 & 0.0017 & 0.01664 & 0.00016 & 3.084 & 0.117 \\
2458761.876793 & 17407.12 & 2.16 & 3.37341 & 15.93166 & -0.00303 & 0.08387 & 0.00024 & 0.07517 & 0.00058 & 0.1466 & 0.0019 & 0.01724 & 0.00018 & 3.192 & 0.134 \\
2458763.557347 & 17424.26 & 1.61 & 3.38776 & 16.18228 & 0.00645 & 0.08525 & 0.00019 & 0.07743 & 0.00041 & 0.1530 & 0.0012 & 0.01674 & 0.00013 & 3.265 & 0.054 \\
2458763.568863 & 17424.55 & 1.54 & 3.38635 & 16.16532 & 0.00706 & 0.08374 & 0.00019 & 0.07494 & 0.00038 & 0.1493 & 0.0011 & 0.01656 & 0.00012 & 3.208 & 0.049 \\
2458764.859828 & 17417.30 & 1.47 & 3.35755 & 15.94340 & 0.00174 & 0.08793 & 0.00017 & 0.08263 & 0.00042 & 0.1666 & 0.0014 & 0.01736 & 0.00011 & 3.335 & 0.079 \\
2458764.871344 & 17419.59 & 1.51 & 3.35725 & 15.92168 & -0.00758 & 0.08945 & 0.00018 & 0.08653 & 0.00045 & 0.1739 & 0.0015 & 0.01809 & 0.00012 & 3.400 & 0.083 \\
2458766.552229 & 17421.07 & 1.77 & 3.39656 & 16.29566 & -0.00427 & 0.08907 & 0.00022 & 0.08453 & 0.00046 & 0.1610 & 0.0013 & 0.01795 & 0.00015 & 3.310 & 0.054 \\
2458766.563548 & 17417.80 & 1.67 & 3.40127 & 16.27750 & -0.00614 & 0.08924 & 0.00021 & 0.08587 & 0.00044 & 0.1655 & 0.0012 & 0.01800 & 0.00014 & 3.419 & 0.050 \\
2458773.576203 & 17406.84 & 4.44 & 3.40203 & 16.14521 & 0.00887 & 0.08427 & 0.00052 & 0.07264 & 0.00103 & 0.1497 & 0.0032 & 0.01696 & 0.00045 & 2.869 & 0.190 \\
2458773.692496 & 17404.76 & 2.58 & 3.38599 & 16.13724 & 0.00387 & 0.08447 & 0.00030 & 0.07690 & 0.00063 & 0.1823 & 0.0021 & 0.01717 & 0.00023 & 3.116 & 0.109 \\
2458774.661052 & 17419.05 & 1.42 & 3.38376 & 16.29187 & -0.00735 & 0.08296 & 0.00018 & 0.07487 & 0.00035 & 0.1492 & 0.0010 & 0.01644 & 0.00011 & 3.504 & 0.037 \\
2458774.753132 & 17420.22 & 1.74 & 3.37372 & 16.05479 & -0.00494 & 0.08526 & 0.00021 & 0.08016 & 0.00046 & 0.1614 & 0.0014 & 0.01723 & 0.00014 & 3.747 & 0.070 \\
2458776.768016 & 17423.26 & 4.61 & 3.38822 & 15.96856 & -0.02925 & 0.08095 & 0.00051 & 0.06867 & 0.00112 & 0.1496 & 0.0039 & 0.01664 & 0.00046 & 2.744 & 0.254 \\
2458776.835004 & 17422.78 & 4.87 & 3.37391 & 15.89282 & 0.01259 & 0.08212 & 0.00053 & 0.07226 & 0.00121 & 0.1523 & 0.0043 & 0.01630 & 0.00050 & 2.740 & 0.302 \\
2458777.586320 & 17425.70 & 3.31 & 3.38196 & 16.17172 & -0.00043 & 0.08125 & 0.00037 & 0.07014 & 0.00079 & 0.1439 & 0.0025 & 0.01674 & 0.00031 & 3.038 & 0.153 \\
2458777.723363 & 17422.05 & 2.52 & 3.38064 & 16.13028 & 0.00371 & 0.08470 & 0.00030 & 0.07866 & 0.00063 & 0.1585 & 0.0020 & 0.01741 & 0.00022 & 3.271 & 0.115 \\
\hline
\end{tabular}
\tablefoot{The table lists the RV, CCF full width at half maximum (FWHM), contrast, and bisector span, H$\alpha$, H$\beta$, H$\gamma$, NaD, and S$_{\rm HK}$ indexes.}
\end{table*}

\begin{table}
\small
    \renewcommand{\arraystretch}{1.25}
    \setlength{\tabcolsep}{2pt}
\centering
\caption{Modelling of the SED.}\label{table.sed}
\begin{tabular}{lccc}
\hline
\hline
Parameter & & Prior & Posterior median   \\
&  & & and 68.3\% CI \\
\hline
Effective temperature, $T_{\mathrm{eff}}$ & [K]     & $N$(3556, 70)     & 3659$^{+50}_{-63}$ \\
Surface gravity, \logg\                   & [cgs]   & $U$(-0.5, 6.0)    & 5.33$^{+0.44}_{-0.56}$ \\
Metallicity, $[\rm{Fe/H}]$                & [dex]   & $N$(-0.06, 0.12)  & $-0.09 \pm 0.11$ \\
Distance                                  & [pc]    & $N$(22.444, 0.013)& $22.444 \pm 0.013$ \\
$E_{\mathrm{(B-V)}}$                      & [mag]   & $U$(0, 3)         & 0.028$^{+0.042}_{-0.021}$ \\
Jitter Gaia                               & [mag]   & $U$(0, 1)         & 0.107$^{+0.24}_{-0.065}$ \\
Jitter 2MASS                              & [mag]   & $U$(0, 1)         & 0.032$^{+0.061}_{-0.023}$ \\
Jitter WISE                               & [mag]   & $U$(0, 1)         & 0.120$^{+0.092}_{-0.045}$ \\
Radius, $R_\star$                         & [\Rsun] & $U$(0, 100)       & 0.531$^{+0.016}_{-0.012}$ \\
Luminosity                                & [L$_\odot$] &               & 0.0455$^{+0.0018}_{-0.0016}$ \smallskip\\
\hline
\end{tabular}
\tablefoot{$N$($\mu$,$\sigma$): Normal distribution prior with mean $\mu$, and standard deviation $\sigma$. $U$(l,u): Uniform distribution prior in the range [l, u].}
\end{table}

\begin{figure}
  \centering
  \includegraphics[width=0.49\textwidth]{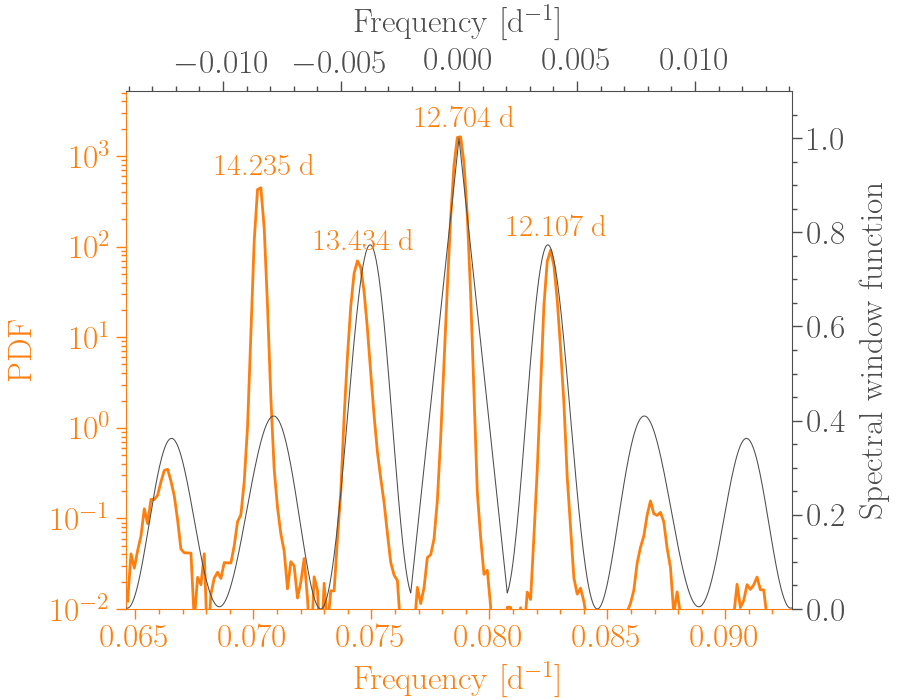}
  \caption{Posterior of the trial frequency for the second planet search (orange, representing the same colour as in Fig.~\ref{fig.search}), with the spectral window function (grey, provided by \protect\url{dace.unige.ch}) centred at the highest peak.}\label{fig.alias}
\end{figure}
\footnotetext{}

\begin{table}[h]
\renewcommand{\arraystretch}{1.25}
    \scriptsize
\caption{Two-Keplerian fit to the HARPS data.}\label{table.RV}
\begin{tabular}{lccc}
\hline
\hline
Parameter & Units &  Prior & Posterior \\
\hline
$\gamma$              &[\kms]    & $U$(17.395, 17.425) & $17.4098 \pm 0.0046$ \\
Jitter                &[\ms]     & $J$(0.01, 5)        & $0.102^{-0.081}_{+0.46}$ \smallskip\\
\emph{GJ~3090~b}\\
P                      &[d]      & $N$(2.8531042, $3.1\times10^{-6}$) & $2.8531042 \pm 3.2\times10^{-6}$  \\
T$_{\rm c} - 2,450,000$&[BJD]    & $N$(8370.41862, 0.00030)           & $8370.41862 \pm 0.00030$ \\
$\sqrt{e}\cos{\omega}$ &         & $U$(-1, 1)    & $-0.25^{-0.16}_{+0.25}$ \\
$\sqrt{e}\sin{\omega}$ &         & $U$(-1, 1)    & $-0.10 \pm 0.36$ \\
e                      &         &               & $0.18^{-0.12}_{+0.14}$ \\
$\omega$               &[\degree]&               & $207^{-78}_{+51}$ \\
K                      &[\ms]    & $U$(0, 10)    & $2.39 \pm 0.56$  \smallskip\\
\emph{GJ~3090~c}\\
P                      &[d]      & $U$(12.5, 12.9)& $12.734^{-0.033}_{+0.026}$ \\
T$_{\rm c} - 2,450,000$&[BJD]    & $U$(8364, 8377)& $8370.52^{-0.97}_{+1.4}$  \\
$\sqrt{e}\cos{\omega}$ &         & $U$(-1, 1)    & $0.40^{-0.30}_{+0.15}$\\
$\sqrt{e}\sin{\omega}$ &         & $U$(-1, 1)    & $-0.01 \pm 0.23$\\
e                      &         &               & $0.22 \pm 0.13$ \\
$\omega$               &[\degree]&               & $-1^{-43}_{+36}$ \\
K                      &[\ms]    & $U$(0, 20)    & $6.18 \pm 0.76$  \smallskip\\
\emph{QPC}\\
$h_1$                  & [\ms]   & $J$(0.01, 50) & $9.6^{-2.3}_{+3.1}$ \\
$h_2$                  & [\ms]   & $J$(0.01, 50) & $0.49^{-0.46}_{+4.0}$ \\
$P$                    & [d]     & $U$(1, 30)    & $17.733^{-0.039}_{+0.21}$ \\
$\lambda$              & [d]     & $U$(1, 2000)  & $610^{-350}_{+450}$ \smallskip\\
\hline
\end{tabular}
\tablefoot{$N(\mu, \sigma)$: Normal distribution with mean $\mu$, and standard deviation $\sigma$. $U(a, b)$: A uniform distribution defined between a lower $a$ and upper $b$ limit. $J(a, b)$: Jeffreys (or log-uniform) distribution defined between a lower $a$ and upper $b$ limit. $\gamma$, P, T$_{\rm c}$, e, $\omega$, and K denote the systemic velocity, orbital period, time of conjunction, eccentricity, argument of pericentre, and RV semi-amplitude, respectively. $h_1$, $h_2$, $P$, and $\lambda$ are the hyperparameters of the QPC kernel defined in \citet{perger2021}.}
\end{table}

\begin{figure*}
\centering
\includegraphics[width=1.0\hsize]{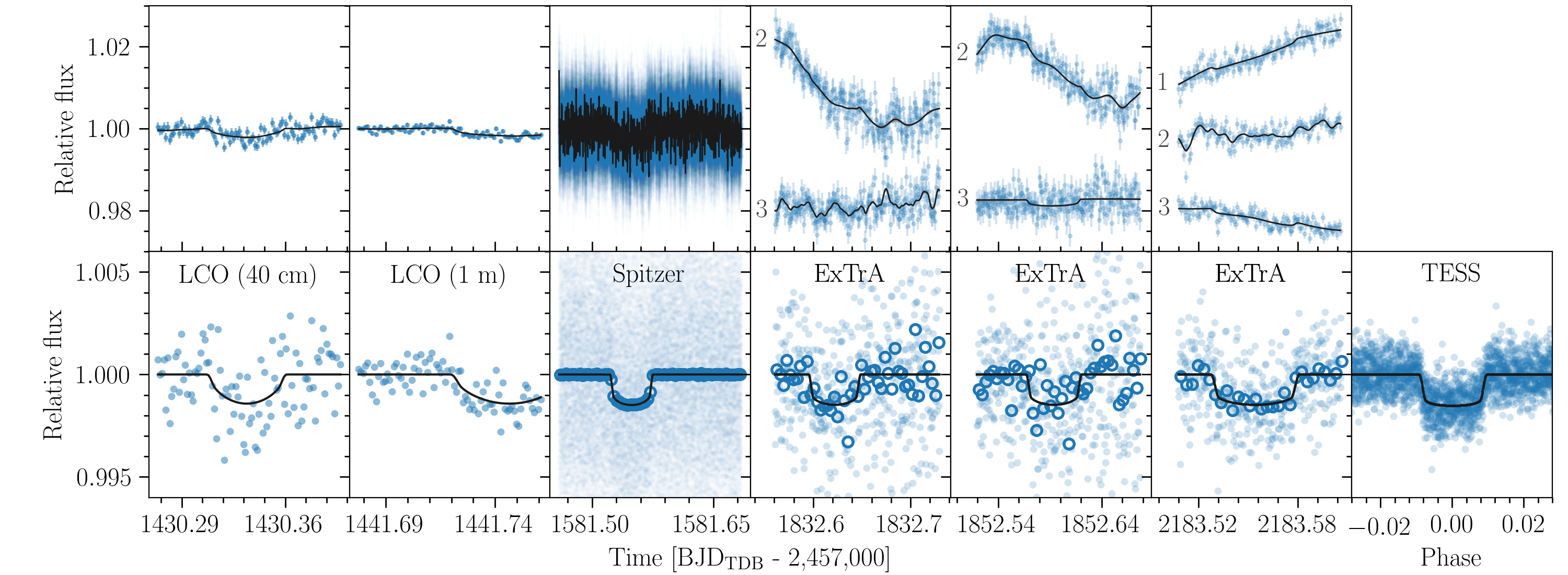}
\includegraphics[width=1.0\hsize]{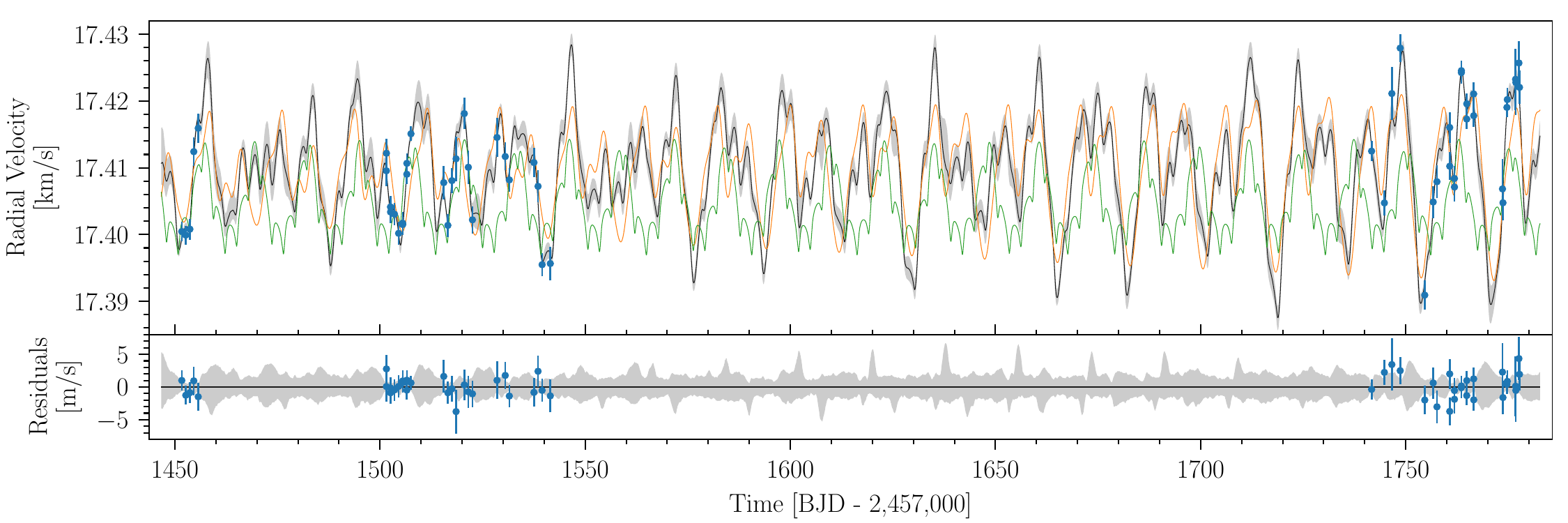}
\includegraphics[width=0.497\hsize]{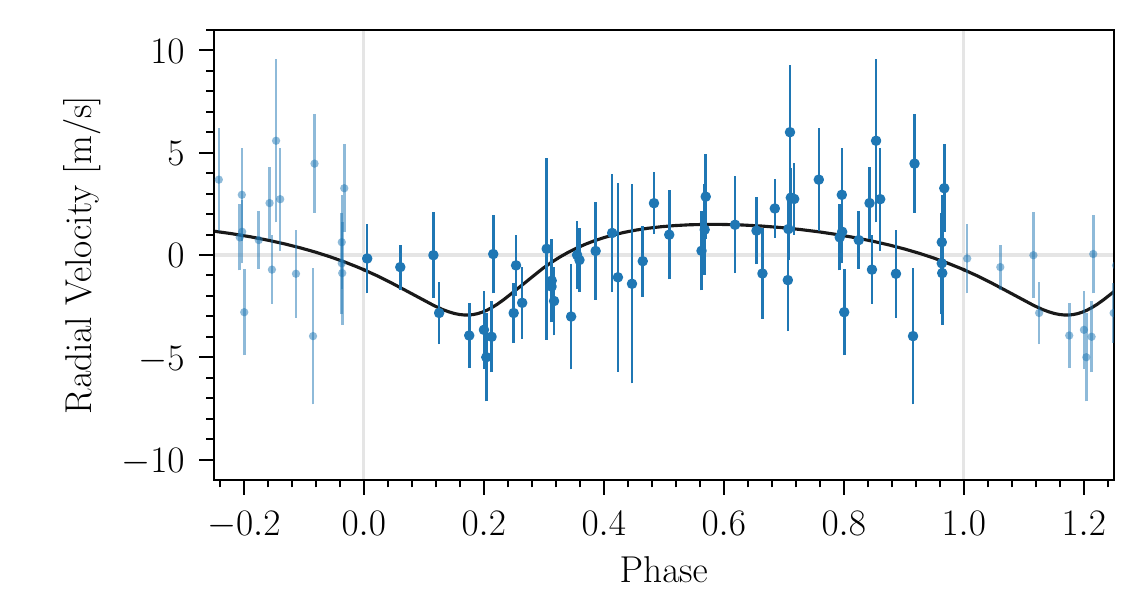}
\includegraphics[width=0.497\hsize]{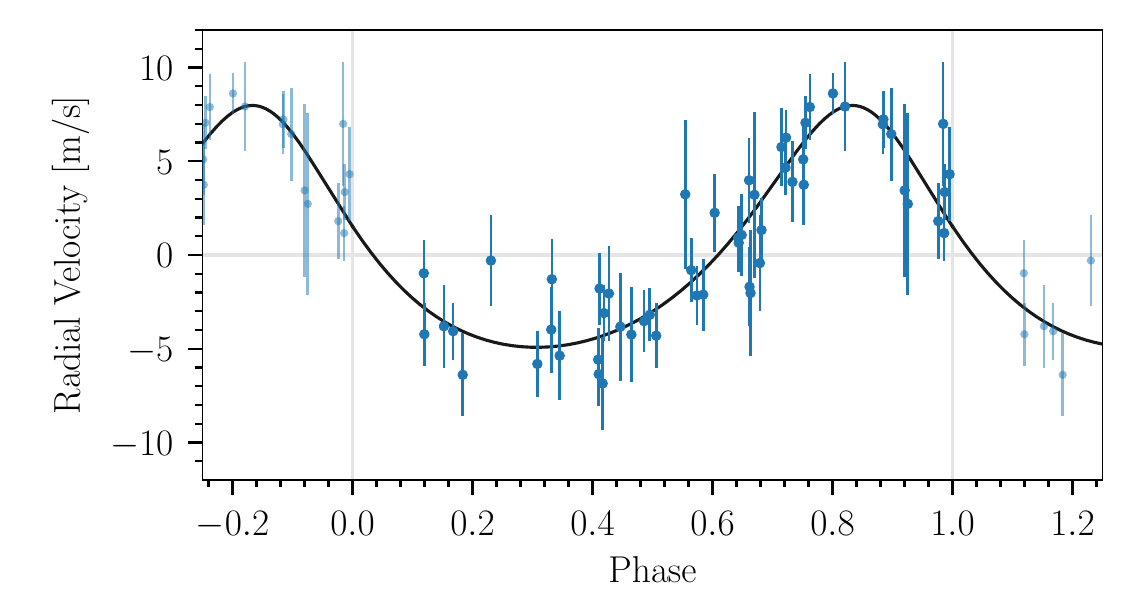}
\caption{Keplerian modelling of transit photometry and RVs of GJ~3090. {\it Top panel}: Transit photometry observations (top, blue error bars), median model (black line), and 68\% CI (grey band, barely visible) computed from 1000 random posterior samples. For the ExTrA observations, the label numbers (1, 2, and 3) correspond to each of the ExTrA telescopes (offset for clarity). Data (bottom, blue dots) detrended with the MAP model (black line). For the Spitzer and ExTrA observations, 5-minute bins are shown as blue open circles. {\it Middle panel}: HARPS RVs (top, blue error bars, as computed from the spectra), median model (black line), 68\% CI (grey band), two-Keplerian MAP model (green line), and the mean of the predictive distribution of the kernel for the MAP model (orange line). Residuals to the median model (bottom). {\it Bottom panel}: HARPS RVs (bottom, blue error bars) corrected for the MAP model of activity (the orange line in the middle panel) and the Keplerian orbit of planet~c, folded at the period of planet~b. The black line is the corresponding Keplerian model of GJ~3090~b with the parameters of the MAP. The same is shown for planet~c in the right panel.}\label{fig.juliet}
\end{figure*}

\begin{figure*}
  \hspace{-2cm}\includegraphics[width=1.2\textwidth]{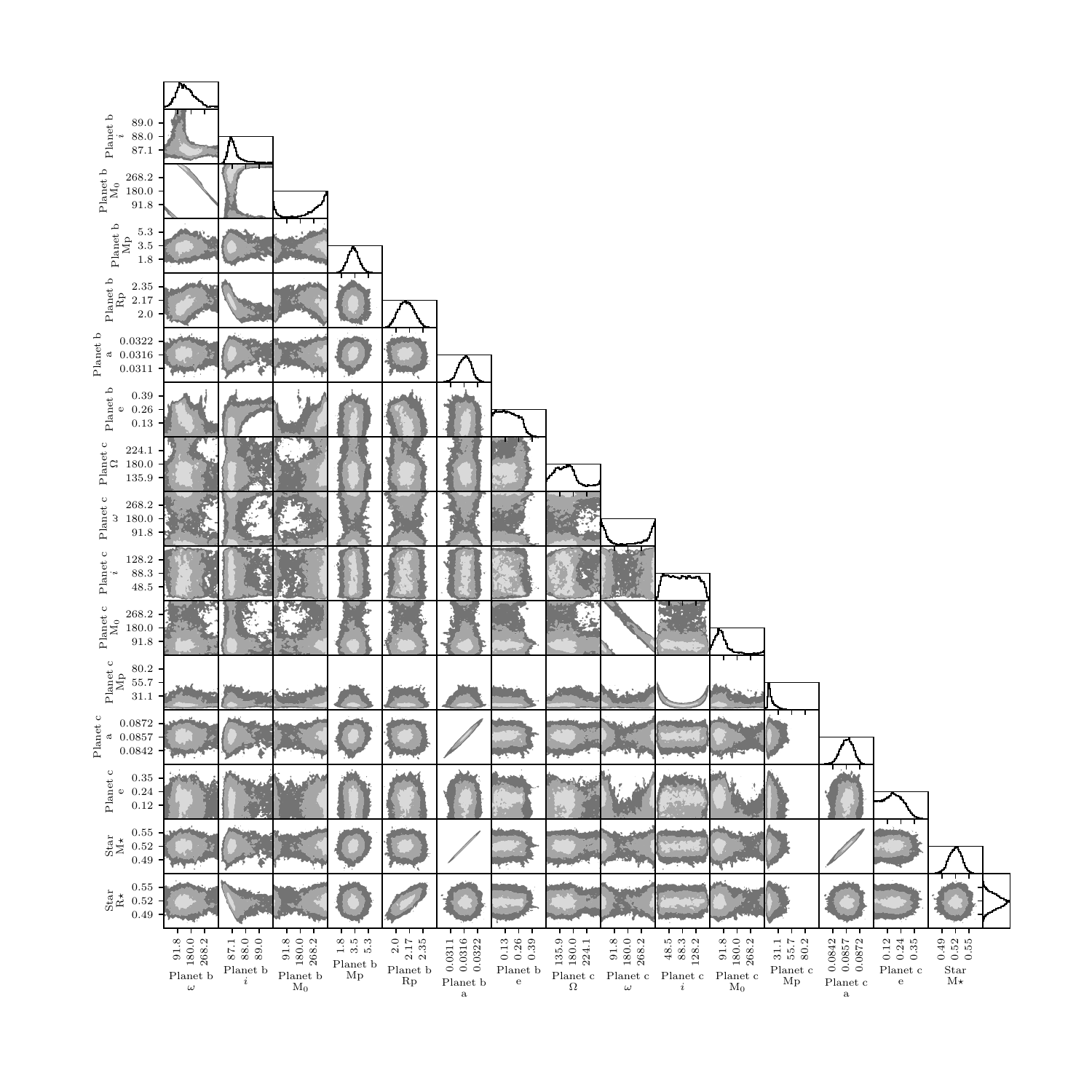}
  \vspace{-2cm}\caption{Two-parameter joint posterior distributions for the most relevant model parameters of the photodynamical modelling with the stability constrain. The 39.3, 86.5, and 98.9\% two-variable joint confidence regions are denoted by three different grey levels. In the case of a Gaussian posterior, these regions project on to the one-dimensional 1, 2, and 3~$\sigma$ intervals. The histogram of the marginal distribution for each parameter is shown at the top of each column, except for the parameter on the last line, which is shown at the end of the line. Units are the same as in Table~\ref{table.results}.} \label{fig.pyramid}
\end{figure*}

\begin{figure}
  \centering
  \includegraphics[width=0.49\textwidth]{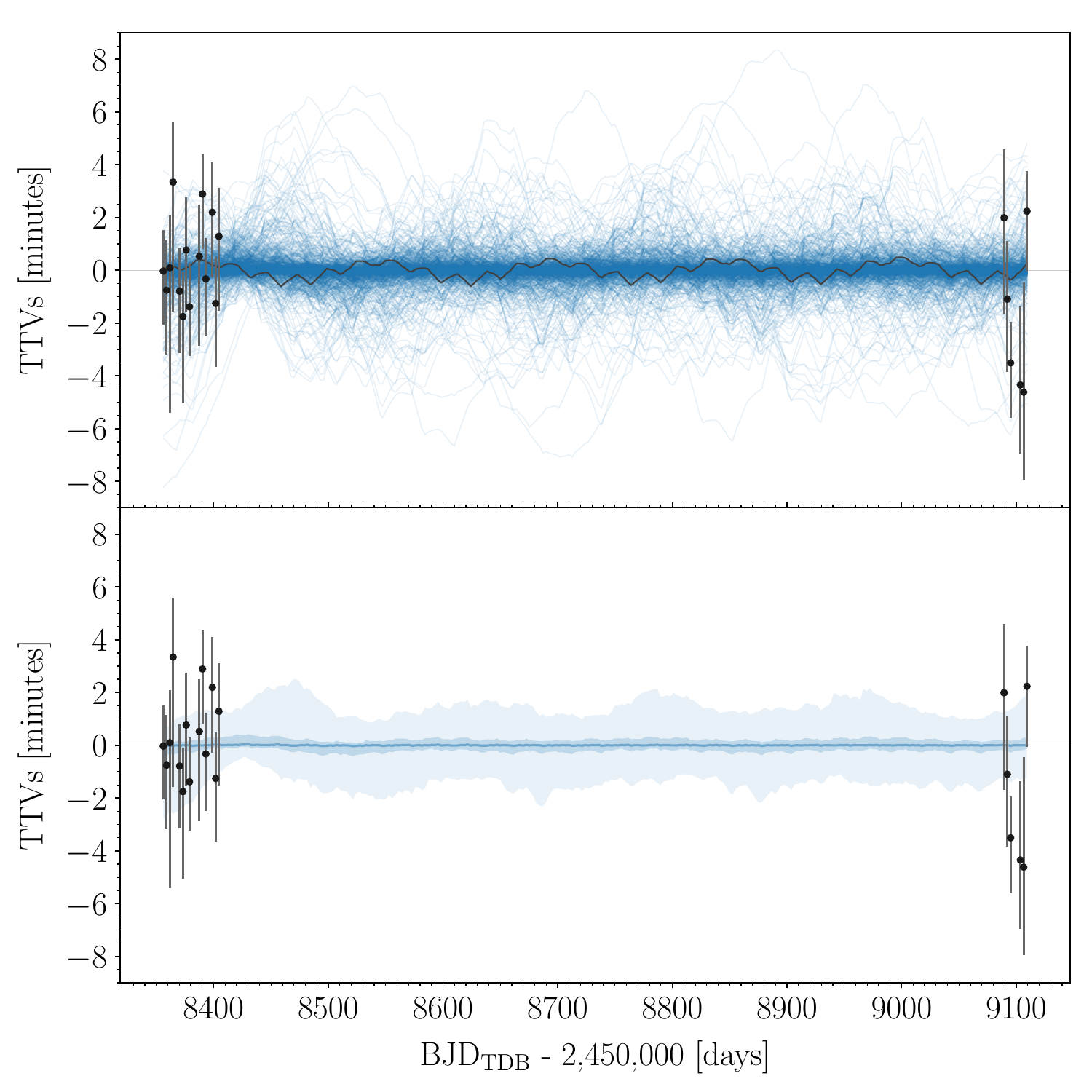}
  \caption{Posterior TTVs of GJ~3090~b from the photodynamical modelling (blue) compared with the with individual transit time determinations (black error bars). {\it Top panel}: One thousand random draws from the posterior distribution are shown. To compute the TTVs, we used a different linear ephemeris for each posterior sample and the period and time of conjunction from Sect.~\ref{sec.juliet} for the individual transit time determinations. {\it Lower panel}: The 68.3\%, and 95.4\% CIs are plotted in different intensities of blue.} \label{fig.TTVs}
\end{figure}

\end{appendix}
\end{document}